\documentclass[prd,%
preprint,%
tightenlines,superscriptaddress,showpacs,%
nofootinbib,eqsecnum,amsfonts,amsmath]{revtex4}
\usepackage{bm}
\newcommand{\alphaa}{\mbox{\boldmath $\alpha$}}
\newcommand{\betaa}{\mbox{\boldmath $\beta$}}

\newcommand{\e}{\mbox{$O(\epsilon )$}}
\newcommand{\ee}{\mbox{$O(\epsilon^2 )$}}

\newcommand{\de}{\mbox{$O(\delta_{\epsilon} )$}}
\newcommand{\me}{\mbox{${\cal M}^{\epsilon}$}}
\begin{document}
\title{The graviton vacuum as a distributional state in kinematic Loop 
Quantum Gravity}


\author{Madhavan Varadarajan}\email{madhavan@rri.res.in} 
\affiliation{Raman Research
Institute, Bangalore 560 080, India}
\affiliation{Sektion Physik, LMU, Theresienstr.37, Munich,  Germany}

\begin{abstract}
The quantum behaviour of weak gravitational fields admits an adequate, albeit
approximate, description by those graviton states in which the expectation values and
fluctuations of the linearised
gravitational field are small. Such states must approximate 
corresponding states in full quantum
gravity. We analyse the nature of this approximation for the 
graviton vacuum state in the context of kinematical Loop Quantum Gravity (LQG)
wherein the constraints are ignored. 
We identify the graviton vacuum state with  kinematically non-normalizable,
distributional states in LQG 
by demanding that relations between linearised operator actions
on the former are mirrored by those of their non-linear counterparts on 
the latter. We define a semi- norm on the space of kinematical distributions
and show that the identification is approximate upto distributions 
which are small in this semi-norm.
We argue that our candidate states are annihilated 
by the linearised constraints
(expressed as operators in the full theory) to 
leading order in the  parameter characterising the approximation.
This suggests the possibility, in a scheme such as ours, of
solving the full constraints order by order in this parameter. The main 
drawback of our considerations is that they depend on certain auxilliary
constructions which, though mathematically well defined, do not
arise from  physical insight.
Our work is an attempt to implement an earlier proposal of Iwasaki and 
Rovelli.

\end{abstract}
\maketitle

\widetext
\baselineskip22pt

\section{Introduction}

A Dirac constraint quantization of a Hamiltonian formulation of gravity is 
defined through the following steps. First, a ``kinematical'' representation
of the Poisson bracket algebra of a large enough set of functions on the
unconstrained phase space is constructed such that these functions act
as linear operators on the representation space. Next, the constraints of
the theory are represented as quantum operators in this representation
and physical states are identified with their kernel. Finally, an inner product
on the space of physical states is chosen
which enforces hermiticity conditions on a complete set of operators 
corresponding to Dirac observables, thus converting the physical state space
to a Hilbert space. Since one expects the Dirac observables to have direct
physical interpretation, the physical interpretation of the formalism
follows.\footnote{We ignore here, the thorny issues of the applicability of
the Copenhagen interpretation to quantum gravity.} Within the Loop Quantum
Gravity (LQG) approach, the kinematical structures are well understood. 
There has also been significant progress in finding physical states.
Further progress is obstructed by the fact that almost no Dirac observables
are known for general relativity. Indeed, this is a problem of the 
{\em classical} theory. As in any generally covariant theory, 
evolution in general relativity   
(at least in the spatially compact case) is 
generated by the constraints, thus implying that Dirac observables are 
constants of motion. Hence finding enough Dirac observables is equivalent to 
finding a complete set of constants of motion for the Einstein equations,
a task which is well nigh impossible. Therefore, to achieve progress in
interpreting the formalism, ideas are needed as to how to obtain a physical 
interpretation  of physical states in the absence of explicit Dirac 
observables.

In contrast to this state of affairs for full blown general relativity, an 
adequate quantum description of  weak gravitational fields {\em is} available.
Sufficiently weak gravitational fields are described by a linearization of  
general relativity about flat spacetime. Quantization of this (approximate)
description of the exact physics is in terms of the graviton Fock 
representation. Since the exact description is that of full quantum gravity,
Fock states in linearised gravity which describe weak gravitational fields
must correspond, at least approximately, to states in
full quantum gravity.
Note that only those Fock states in which the gravitational field has small 
expectation values and fluctuations describe weak `quantum' gravitational 
fields and it is only such states which are expected to have full quantum 
gravity correspondents. An archetypal example of such a state is the graviton
vacuum which is expected to provide an adequate physical description of small 
quantum gravitational fluctuations about flat spacetime. 

Let us suppose that within the LQG framework, we find a way to associate
{\em kinematical} 
structures with Fock states  describing weak gravitational 
fields in the context of
a mathematically well defined approximation scheme. Since kinematical 
structures play a key role in defining physical states, we may hope that the 
nature of the approximation scheme suggests how to extend this correspondence
(of appropriate Fock states with kinematical structures) to physical states.
Such a correspondence would provide a way to interpret these  physical
states (as small quantum perturbations of Minkowski spacetime) in the 
absence of Dirac observables.

Motivated  by these remarks, our aim in this work is to make a preliminary 
investigation of the relation  between quantum linearised gravity  and
LQG by finding structures in kinematical LQG which correspond to the 
Fock vacuum. As we shall see, a reasonable definition of such a correspondence
requires the satisfaction of infinitely many conditions by the putative 
LQG structures and therefore the identification of these structures 
is quite involved. 

We proceed as follows.
In section 2, we analyse 
the conditions under which classical linearised 
gravity is an adequate approximation to full general relativity 
and show that
the approximation may be characterised by a small parameter, $\epsilon$,
constructed out of the physically relevant classical distance scales.
\footnote{ We are indebted to Abhay Ashtekar for his crucial inputs without which
the analysis and results of  section 2 would not have been possible.}
Thus, the equations of motion of classical linearised theory are 
obtained by an expansion of the equations of motion of full general
relativity about flat spacetime in which only terms of linear order in 
$\epsilon$ are retained.
For the linearised theory vacuum to be of physical relevance, it is 
necessary that vacuum expectation values and fluctuations of the gravitational
field be of order $\epsilon$. \footnote{ We only expect this condition to hold
when the vacuum is probed at `large distance scales. A detailed  discussion
of these matters is given in section 2}.
We analyse this requirement in the context of the $r$- Fock representation
of linearised gravity \cite{ars,meem,melingrav} 
and show that it leads 
to a specification of $\epsilon$ in terms of 
an intertwining of the 
physically relevant classical distance scales with the distance scales
relevant to the quantum theory, namely the Planck length and the scale $r$
characterising the $r$- Fock representation. We use the $r$- Fock 
representation because of its structural similarity to the representation 
used for LQG.  The $r$- Fock representation of linearised gravity is
constructed in \cite{ars,melingrav} (see also Appendix 2 of \cite{carlo}) 
and its 
relation to the usual Fock representation is discussed in \cite{melingrav}.

In section 3, we propose to identify states in kinematical 
LQG  with the $r$- Fock vacuum by demanding that  the action of the 
linearised theory operators  (to order $\epsilon$) 
on the $r$- Fock vacuum be mirrored by the 
action of their non-linear counterparts in LQG. We show that this demand
leads to infinitely many consistency conditions and the consequent 
necessity of identifying an infinite dimensional set of Fock states with 
their counterparts in LQG.  We also argue that these counterparts 
are kinematical {\em distributions} i.e. they are kinematically 
non-normalizable states which lie in the algebraic dual to the finite
span of  extended spin network states (for a definition of the latter
see \cite{area}). We characterise this identification by a map ${\cal M}$
\footnote{We actually use a {\em 1 parameter family of maps}, ${\cal M}^{\epsilon}$,
 in section 3. For pedagogical reasons we gloss over this subtleity
and use notation appropriate to a single map  
${\cal M}$ in this section.}
from such distributions to appropriate subspaces of $r$-Fock space.
The material after section 3 is devoted to showing the existence of
at least one such map which satisfies the consistency conditions.

In section 4 we define two auxilliary structures in LQG which go into 
our construction of $\cal{M}$. The first is a distributional `weave' state 
which appropriately captures the notion of classical flat space used in the 
linearised theory. The second is a semi-norm on the space of distributions.
The kernel of $\cal{M}$ is (partially) specified  by those distributions
which are of order $\delta_{\epsilon}$ in this semi- norm, where $\delta_{\epsilon}$
is a small $\epsilon$- dependent parameter. 
In this section we also specify the 
general 
structure of our putative solutions to the consistency conditions.
In section 5 we construct such solutions explicitly thus exhibiting 
distributional states in LQG which correspond to the $r$- Fock vacuum.
In section 6 we argue
 that the constraints of the linearised theory, represented
as operators in the full theory, 
map the 
distributional states we constructed in section 5 to the kernel of  $\cal M$
i.e. their images are of 
order $\delta_{\epsilon}$. 
Section 8 is devoted to a discussion of our results and a few important
identities are proved in the Appendices.

Although our work
suggests a strategy to solve the constraints of full gravity order by
order,
it is pertinent to make a couple of  cautionary remarks here to 
place our results in  proper perspective.\\
\noindent (a) Our results would be stronger if we could prove them
in the context of a norm rather than a semi- norm. In fact, most of the considerations
of sections 5 and 6 {\em do} go through in the context of a suitably defined norm.
The only failure in this context is that of our states to be mapped into the 
kernel of ${\cal M}$ by the linearised Gauss law constraint. In section 7 
we comment on this
in relation to the fact that the connection variables of linearised theory 
and LQG take values in distinct  Lie algebras (namely $U(1)^3$ and $SU(2)$). \\
\noindent (b) As noted earlier the $r$- Fock  construction is structurally similar to
LQG. 
As briefly discussed in section 7
our constructions of $\cal M$, the semi- norm as well as the distribution 
corresponding to the vacuum state, all  lean
heavily on the existence of similar structures in the $r$- Fock representation.
This is the reason that our constructions `work' at a mathematical level.
However, they are still {\em ad hoc} because they do not arise from 
any deep {\em physical} insight. 

In view of (a) and (b)  above, it is  unlikely that the detailed
choices of auxilliary structures made in this work lead to a direct relation
between ``solutions to the constraints to $n$th order''
and exact physical states in nonperturbative LQG. Neverthless, our hope is that
further analysis may lead to a more physically motivated implementation of
the general strategy (described in section 3) suggested by our work. 
If so, this would provide 
 a plausible  interpretation of certain states in LQG (namely those corresponding
to the vacuum of linearised gravity) without recourse to Dirac observables.

There have been other attempts to identify the vacuum state in LQG 
\cite{ttvac,aavac,ttcomplexifier,conrady} as well as to understand how {\em classical}
spacetime emerges fron LQG structures \cite{weave,statgeom,ttvac}.
Indeed, these questions are the focus
of alot of current research in the field. The related problem of interpretation
of physical states in LQG is a very crucial and hard problem. Though the 
current set of ideas (including ours) may be critiscized on several fronts,
we believe that only an exploration of a diversity of such proposals
will suggest some way to attack this problem.

Finally, we note that this work is an attempt to implement earlier proposals
of Iwasaki and Rovelli \cite{IR}.
\footnote{Note that J. Zegwaard independently explored the same issues as 
Iwasaki and Rovelli using a slightly different approach in his work \cite{joost}.}
Indeed, this work
would not have been possible without 
their considerations. It is our view that their beautiful ideas were 
considerably in advance of the technical state of art of the field and 
consequently were difficult to implement with precision. We hope that 
we have partially remedied this by our work here.

In what follows we shall assume familiarity with the loop representation of
linearised gravity  as discussed in 
\cite{melingrav}. We set $c=G=1$ and denote the Planck length by $l_P$.
We also define $O(\epsilon^n )$ and $O^{\infty}(\epsilon )$ as follows. Let $x$ be an 
$\epsilon$- dependent complex number. $x=O(\epsilon^n )$ iff 
$\lim_{\epsilon \rightarrow 0} {|x|\over \epsilon^n}$ exists. $x=O^{\infty}(\epsilon )$ iff
as $\epsilon \rightarrow 0$, $x\rightarrow 0$ faster than any positive power of $\epsilon$. 

\section{Linearised gravity as an approximate physical description.}

\subsection{A quick review of linearised gravity in terms of connections.}

The classical Hamiltonian formulation underlying LQG is discussed in 
\cite{realsu2}.The phase space variables are a spatial 
$SU(2)$ connection, $A_a^i ({\vec x})$ and a densitized triad field 
$E^b_j({\vec y})$. 
Here
$a,b$ denote spatial components, $i,j$ denote internal $SU(2)$ Lie algebra
components.
To define the linearised theory about a flat background, the spatial toplogy
is chosen to be $R^3$. We
fix, once and for all, a cartesian coordinate system $\{ {\vec x}\}$ on $R^3$
as well as an orthonormal basis in the Lie algebra of $SU(2)$. Henceforth
all components  refer to  this cartesian coordinate system 
and to this internal basis. We linearise the $SU(2)$ formulation about the 
phase space point $(A_a^i=0, E^a_i = \delta^a_i)$. As in \cite{melingrav},
we denote the linearised triad field by $e^a_i$  so that 
\begin{equation}
E^a_i = \delta^a_i + e^a_i .
\label{ltriad}
\end{equation}
Since the background connection vanishes, there is no need to introduce a new
symbol for the linearised connection. The only non-vanishing
Poisson bracket is
\begin{equation}
\{A_a^i({\vec x}), e^b_j({\vec y})\}_L = {\gamma} 
         \delta^b_a \delta^i_j 
               \delta ({\vec x}, {\vec y}).
\label{lpb}
\end{equation}
and the linearised Gauss Law, Vector and Scalar constraints are
\begin{eqnarray}
G^L_i &=& \partial_a e^a_i + \epsilon_i^{\;ja}A_{aj} , 
\label{lgauss}\\
V^L_a &=& f_{ab}^a,
\label{lvector} \\
C^L &=& \epsilon^{abc} f_{abc}. 
\label{lscalar}
\end{eqnarray}
Here $\gamma$ is the Barbero-Immirzi parameter, $f_{ab}^i = \partial_a A_b^i-\partial_b A_a^i$ is the linearised
curvature, $\partial_a$ denotes the flat derivative operator which 
annihilates the background triad and we use the
background triad $\delta^a_i$ to freely interchange internal 
and spatial indices. The linearised $SU(2)$ connection transforms 
as a triplet of $U(1)$ connections under transformations generated
by $G^i_L$.  
  
Given any oriented piecewise analytic loop $\beta$ in $R^3$ with (arbitrary)
parametrization $s$ and tangent vector ${\dot{\beta}}^a$, we define the 
loop form factor
\begin{equation}
X^a_{\beta}({\vec x}):=
\oint_{\beta} ds \delta^3({\vec {\beta}} (s), {\vec x}){\dot{\beta}}^a,
\label{eq:x}                                
\end{equation}
and its Fourier transform
\begin{equation}
X^b_{\beta}({\vec k})={1\over (2\pi)^{3\over 2}}\int d^3x
X^b_{\beta}({\vec x}) e^{-i{\vec k}\cdot {\vec x}}.
= {1\over (2\pi)^{3\over 2}}\oint_{\beta} ds e^{-i{\vec k}\cdot {\vec \beta}(s)}{\dot{\beta}}^a
\label{xk}
\end{equation}
For any positive parameter $r$ with dimensions of length, we define the Gaussian 
smeared loop form factor 
\begin{eqnarray}
X^a_{\beta (r)}({\vec x}) &:=& \int d^3y X^a_{\beta} 
({\vec y}){e^{-{|{\vec x}- {\vec y}|^2 \over 2 r^2}}\over (2\pi r^2)^{3\over 2}} \\
X^a_{\beta (r)} ({\vec k}) 
&=& e^{-{k^2r^2\over 2}} X^a_{\beta} ({\vec k}) .
\label{xrk}
\end{eqnarray}
For any triplet of loops $\alpha_i, i=1..3$ we define
$X^{ab}_{\alphaa}
({\vec x})$ and 
$X^{ab}_{\alphaa (r)}({\vec x})$ via their  Fourier transforms to be,
\begin{equation}
X^{ab}_{\alphaa}({\vec k}):=
X^a_{\alpha_i}({\vec k})\delta^{ib} , \;\;\;
X^{ab}_{\alphaa (r)}({\vec k}) := e^{-{k^2r^2\over 2}} X^a_{\alphaa} ({\vec k}).
\label{xab}
\end{equation}
We also define  $G^{ab}_{\alphaa(r)}({\vec x})$ via its Fourier transform,
\begin{equation}
G^{ab}_{\alphaa(r)}({\vec k}):=
 k X^+_{\alphaa (r)}({\vec k}) (1+2i\gamma )m_am_b
- k X^-_{\alphaa (r)}({\vec k}) (1-2i\gamma ){\bar m}_a{\bar m}_b .
\label{gab}
\end{equation}
Here $m_a,{\bar m}_a$ form the standard transverse basis in momentum space
and $X^{\pm}_{\alphaa (r)}$ are the positive and negative 
helicity components of the transverse, traceless, symmetric part of 
$X^{ab}_{\alphaa (r)}$. Finally we define the following Dirac 
observables for the linearised theory:
\begin{equation}
h_{\alphaa} = \exp i\int d^3x X^{ab}_{\alphaa}A_{ab}
= \prod_{k=1}^3 \exp i \oint_{\alpha^k}A_a^k dx^a
\label{h}
\end{equation}
and
\begin{equation}
g_{\alphaa (r)} = \int d^3x G^{ab}_{\alphaa (r)} e_{(r)ab},
\label{g}
\end{equation}
where $e_{(r)ab}$ is obtained from $e_{ab}$ by Gaussian smearing.
$h_{\alphaa}$ and $g_{\alphaa (r)}$ play a key role in the 
considerations of \cite{melingrav} and will continue
to play a key role in this work.

\subsection{Classical linearised gravity as an approximate physical description.}

Linearised gravity is expected to provide an approximate description of the physics of
weak gravitational fields. Such fields are characterised by low curvature. 
Since curvature is dimensionful, a notion of its smallness requires the introduction of some parameter
$s_{curv}$ which has dimensions of length. The restriction to small curvatures 
implies that $|\partial_a \partial_b e^c_i | < s_{curv}^{-2}$. On dimensional grounds
we expect phase space data for low curvatures to satisfy 
\begin{equation}
|\partial_a e^b_i|, |A_b^i| < s_{curv}^{-1} .
\label{restrict}
\end{equation}
Clearly, the larger $s_{curv}$ is,  the better is the physical description provided by linearised
gravity.

However, small curvatures are only a necessary condition for the application of linearised gravity.
There exist nonlinear configurations of the gravitational field, namely large black holes, which have 
vast regions of small curvature but also possess global features such as horizons.
Such global features  cannot be described by linearised gravity. More quantitatively,
for data satisfying (\ref{restrict})
we estimate the gravitational energy in a region of size $s$ to be $M= ({1\over s_{curv}})^2 s^3$.
If $M \sim s$ the gravitational configuration is expected to be close to that of a black hole.
Hence, we require that $M <<s$ or, equivalently, that ${s\over s_{curv}} <<1$.
Thus, we expect linearised gravity to be a good description of the gravitational field
when $s_{curv}$ is large and when we probe regions of 
size $s$  much smaller than $s_{curv}$. 
For a fixed probe scale $s$, we define $\epsilon ={s\over s_{curv}}$.
Then linearised gravity is a valid description when $\epsilon <<1$ and equation (\ref{restrict}) reads 
\begin{equation} 
|\partial_a e^b_i|, |A_b^i| <  {\epsilon\over s}.
\label{restrict1}
\end{equation}
In addition, we shall assume that the `observable' part of the linearised triad, namely its positive and
negative helicity parts, also have spatial variations characterised by the scale $s$. Specifically, we assume
\begin{equation}
|\partial_c e^{\pm}_{ab}|,|(-\nabla^2)^{{1\over 2}} e^{\pm}_{ab}|   < {\epsilon\over s} .
\label{restrict2}
\end{equation}
In view of their key role 
 in the quantum theory,
it is useful to encode the conditions (\ref{restrict1}),(\ref{restrict2}) as
conditions on the observables $h_{\alphaa}, g_{\alphaa (r)}$. 
From (\ref{restrict1}) we obtain $|\sum_{k=1}^3 \oint_{\alpha^k}A_a^kdx^a| < {\epsilon\over s} |\alphaa |$,
where $|\alphaa |$ is the sum of the lengths of the loops $\alpha^k , k=1..3$.
In order to get a useful bound on $h_{\alphaa}$, as well as to ensure that the classical probes
(in this case the loops $\alpha^i$) do not extend into regions of volume much larger than $s^3$,
we restrict attention to {\em loops of length at most equal to} $s$. Then 
equations (\ref{h}) and (\ref{restrict1}) imply that 
\begin{equation}
|h_{\alphaa}-1| < 3\epsilon.
\label{hbound}
\end{equation}
With the same restriction on loop lengths, we obtain, using (\ref{xk}) - (\ref{gab}) and (\ref{restrict2})
\begin{eqnarray}
|g_{\alphaa(r)}| &=& |\int X^{ab}_{\alphaa (r)}(\vec{x})
[ (1+2i\gamma ) (-\nabla^2)^{{1\over 2}} e^{+}_{(r)ab}(\vec{x})
+(1-2i\gamma) (-\nabla^2)^{{1\over 2}} e^{+}_{(r)ab}(\vec{x})] |\nonumber\\
& < & 2 \sqrt{1+4\gamma^2} \epsilon .
\label{gbound}
\end{eqnarray}
Note that the above bound is independent of $r$.

\subsection{Quantum considerations}

Equations (\ref{hbound}) and (\ref{gbound}) imply that only those states of quantum 
linearised theory in which the expectation values and fluctuations of the 
operators ${\hat h}_{\alphaa}- 1, {\hat g}_{\alphaa(r)}$ are of order $\epsilon$,
are physically relevant descriptions of quantum aspects of weak gravitational fields.
Since in this work we are interested only in the vacuum state, we require that
the vacuum fluctuations and vacuum expectation values of these operators be 
small (i.e. of $O(\epsilon )$).

As mentioned in the introduction we use the $r$-Fock representation and the $r$-Fock vacuum,
$|0_r>$
to investigate the consequences of this requirement. We denote 
the vacuum expectation value of an operator ${\hat O}$ by 
$\bar O:= <0_r| {\hat O}|0_r>$ and its vacuum fluctuation by 
$\Delta O:= \sqrt{ <0_r| {\hat O}^{\dagger}{\hat O}|0_r> - |{\bar O}|^2}$.
From \cite{melingrav}, we have that ${\bar g_{\alphaa (r)}}= 0$ and that
\begin{eqnarray}
(\Delta g_{\alphaa (r)})^2 &\leq&
{1+4\gamma^2 \over 32\pi^3}l_P^2 l_{\alphaa_{max}}^2 \int d^3k k e^{-k^2r^2} 
\label{2.19a} \\
\leq {1+4\gamma^2 \over 16\pi^2}{l_P^2s^2 \over r^4},
\label{2.19}
\end{eqnarray}
where $l_{\alphaa_{max}}\leq s$ is the length of the longest loop in the triplet $\alphaa$. 
To ensure that $\Delta g_{\alphaa (r)}= O(\epsilon )$  we set,
\begin{equation}
{l_P s\over r^2}= \epsilon .
\label{equant}
\end{equation}
It can be checked that this identication of $\epsilon$ ensures that the vacuum expectation 
values and vacuum fluctuations of ${\hat h}_{\alphaa} -1$ are also at most of  $O(\epsilon )$.
\footnote{ Equation (\ref{2.19}) only provides an upper bound. In appendix E we construct 
$\alpha^k$ such that $\Delta g_{\alphaa (r)}\approx \epsilon$ to nontrivial leading order in 
$\epsilon$.}

How should we interpret equation (\ref{equant})? Note that the parameter $r$ plays no essential role
in classical theory. In particular the considerations of section 2.1 are unaltered if we set $r=0$.
However, for quantum operators to be well defined in a (Fock-like representation of)
a quantum {\em field} theory, it is necessary for the basic quantum fields to be appropriately 
smeared in order to avoid the infinities coming from quantum fluctuations at arbitrarily 
small distance scales.  Here, this smearing is characterised by the distance scale $r$.
While we require that operator fluctuations be small (i.e. of $O(\epsilon )$), the
quantum  uncertainity
principle dictates that operator fluctuations cannot, in general, be arbitrarily small. The 
potential conflict between these two requirements results in the relation (\ref{equant}) 
between the classically relevant distance scale $s_{curv} = {s\over \epsilon}$ and 
the distance scales relevant to quantum theory, namely the Planck length $l_P$ and the 
smearing scale $r$. Equation (\ref{equant}) can be rewritten in the form $r^2 = l_P s_{curv}$.
In this form it is clear that if the smearing scale is too small, the resulting short 
distance quantum fluctuations exceed the upper bounds (\ref{hbound}) and (\ref{gbound}).
Note that we may choose the individual parameters $r,s$ and $s_{curv}$ to vary with 
$\epsilon$ in different ways. Since we would like to have a non-empty set of probes when
$\epsilon \rightarrow 0$, we shall henceforth restrict attention to probe sizes
$s$ which do not vanish as $\epsilon \rightarrow 0$. Then equation 
(\ref{equant}) implies 
that $r$ (and $s_{curv}$) diverge as $\epsilon \rightarrow 0$.
This means that to ensure that the quantum fluctuations in curvature 
become arbitrarily small, the 
relevant operators have to be smeared at arbitrarily large distance scales. 
In this work we shall think
of $r$ as a short distance smearing scale relative to the probe scale $s$ and restrict attention to
the case $s >>r$.

\section{A strategy to identify the vacuum state in LQG.}

In this section we discuss the identification of states in LQG which correspond to the 
$r$-Fock vacuum of linearised theory in terms of a map between (suitably defined)
states in LQG and (subsets of) $r$- Fock space.  Section 3A contains a few pertinent 
observations which serve to motivate the definition and desired properties of this 
map as outlined in section 3B. In section 3C we describe our strategy for an explicit
construction of a map which satisfies the criteria of section 3B.

\subsection{A few remarks by way of motivation.}

\noindent (i){\em Linearised theory observables and their LQG counterparts}:
The operators ${\hat h}_{\alphaa}, {\hat g}_{\alphaa (r)}$ play a key role in the
identification of the $r$-Fock vacuum in linearised theory (see \cite{melingrav}).
Therefore it is reasonable to look for their counterparts in the $SU(2)$ theory.
Using  equation (\ref{restrict1}) and equation (\ref{ltriad}), its is straightforward to 
check that 
\begin{eqnarray}
-i(h_{\alphaa} -1) &=& 2\sum_{k=1}^3 tr H_{\alpha^k} \tau^k  \;\;\; + \;\; O(\epsilon^2 ) ,
\label{hH}\\
g_{\alphaa (r)} &=& G_{\alphaa (r)} .
\label{gG}
\end{eqnarray}
Here $H_{\alpha}$ is the $SU(2)$ holonomy $H_{\alpha}= {1\over 2} P\exp -\oint_{\alpha}A_a^i\tau^i$.
$\tau^i = -{i\over 2} \sigma^i$, where $\sigma^i$ are the Pauli matrices. Thus
$\tau^i$ correspond to the fixed basis of the Lie algebra of $SU(2)$ (see section 2)
in its defining 2 dimensional  representation.
Note that (\ref{hH}) holds independent of which point of $\alpha^k$ is chosen as the base point.
 Also,
\begin{equation}
G_{\alphaa (r)}:=\int d^3x G_{\alphaa (r)\;ab} ({\vec x})(E^{ab}_{(r)} ({\vec x})- \delta^{ab})
\label{GEr}
\end{equation}
$E^{ab}_{(r)}$ is defined from the $SU(2)$ triad variable by 
\begin{equation}
E^{ab}_{(r)} ({\vec x}):= {1\over (2\pi r^2)^{3\over 2}}\int d^3y e^{-{|{\vec x} -{\vec y}|^2\over 2r^2}}
                            E^a_i ({\vec y}) \delta^{bi}.
\label{Eabr}
\end{equation}
We shall refer to the set of functions in linearised theory on the left hand side of equations (\ref{hH}) and (\ref{gG})
as $O^{\epsilon}_L$ and their $SU(2)$ counterparts on the right hand side of these equations as 
$O^{\epsilon}_F$. 
As noted at the end of section 2, $r$ is a function of $\epsilon$
through (\ref{equant}). Since some of observables (namely the ones depending on the triad) depend explicitly on $r$,
we have a superscript $\epsilon$ to signify this dependence. Note also that the linearised theory holonomy operators 
have an implicit dependence on $r$ coming from the $r$- dependence of the representation.

Recall that $O^{\epsilon}_L, O^{\epsilon}_F$ are labelled by loops of size at most $s$. Since we are 
interested in probing regions of size of order $s$ (see section IIB), we further restrict these loops
to lie within a volume $(Ss)^3$ about the origin for some positive parameter {\em $S$ independent of $\epsilon$}.

\noindent (ii) {\em The distributional nature of LQG `vacuum' states}:
It is reasonable to expect that 
any putative vacuum state in the $SU(2)$ theory must be such that the expectation values and fluctuations
of the operators ${\hat O}^{\epsilon}_F$ in this state reproduce the ($r$- Fock) vacuum expectation values and
fluctuations of the operators ${\hat O}^{\epsilon}_L$ to accuracy $\epsilon$.
Since ${\hat O}^{\epsilon}_F$ are not Dirac observables for full gravity, they are only defined at the 
(non $SU(2)$ invariant) kinematic level. Note that the $SU(2)$ states cannot lie in the 
LQG kinematic Hilbert space ${\cal L}^2({\bar {\cal A}}, d\mu_0)$ \cite{al,donze} for the following
reasons. First, as argued in \cite{mejose}, expectation values and fluctuations for uncountably many 
$SU(2)$ holonomy operators cannot be reproduced by states in the kinematic Hilbert space and second,
as can be easily verified, 
 the smeared triad operator (constructed by replacing $E^a_i$ by ${\hat E}^a_i$ in  equation (\ref{Eabr})) 
is not well defined on the kinematic Hilbert space.
As we shall see later, the $SU(2)$ states lie in (an appropriate subspace) of the algebraic dual, $\Phi^*_{kin}$,
to a suitable dense subspace of the kinematic Hilbert space. Since, at least to our knowledge, there is no
natural inner product on $\Phi^*_{kin}$, an identification of the vacuum state therein which does not 
involve explicit evaluation of expectation values and fluctuations of ${\hat O}^{\epsilon}_F$ is desirable.

\noindent (iii) 
{\em The approximate nature of the identification between linearised and LQG structures}:
Finally we note that since linearised gravity is an approximate description of the exact
physics of weak gravitational fields, we expect the correspondence between the $r$-Fock vacuum and
its $SU(2)$ counterparts to be an approximate one.

\subsection{Definition and properties of the map ${\cal M}^{\epsilon}$.}

Motivated by the above remarks, we define the identification of suitable $SU(2)$ distributions 
with the $r$- Fock vacuum, $|0_r>$, as follows. Recall that  
equation (\ref{equant}) implies that $r$ is a function of $\epsilon$. Hence $|0_r>$ corresponds 
to a 1 parameter family of states, 1 state for each $r(\epsilon )$, $0< \epsilon <<1$.
Consider a corresponding 1 parameter family of putative $SU(2)$ vacuum states $\Psi^{\epsilon}_0$.
We require the existence of a 1 parameter family  of maps, ${\cal M}^{\epsilon}$,  which map
the 1 parameter family of states $\Psi^{\epsilon}_0$ to the 1 parameter family of
 {\em sets of} $r-$ Fock states 
$\{|0_r> + |\phi^{\epsilon} >, {\sqrt{<\phi^{\epsilon} |\phi^{\epsilon} >}}= {\ee} \}$.
We denote this requirement as
\begin{equation}
\me \Psi_0^{\epsilon} = |0_r> + \ee .
\label{mphi}
\end{equation}
Thus, for each fixed $\epsilon$,  $\me$ is a map  from a subspace of $\Phi^*_{kin}$ to 
(a subset of) the set of all subsets of $r$- Fock space.
Using similar notation, we require that $\Psi_0^{\epsilon}$ be such that the 
action of any ${\hat O}^{\epsilon}_F$ on $\Psi_0^{\epsilon}$ mirrors that of 
the corresponding ${\hat O}^{\epsilon}_L$
on $|0_r>$:
\begin{equation}
\me {\hat O}^{\epsilon}_F \Psi_0^{\epsilon} = {\hat O}^{\epsilon}_L |0_r> + \ee .
\label{mophi}
\end{equation}
Since we can always measure a linear combination of the basic linearised operators ${\hat O}^{\epsilon}_L $,
we impose the following `linearity' condition on $\me$. Let ${\hat O}^{\epsilon}_{iL},{\hat O}^{\epsilon}_{iF}, i=1..N$
be a set of $N$ linearised and `full' operators and let $a_0, a_i$ be a set of $N+1$ complex parameters each of $O(1)$
with $N$ itself being independent of $\epsilon$. Then we require that 
\begin{equation}
\me (\sum_{i=1}^N a_i {\hat O}^{\epsilon}_{iF} + a_0)\Psi_0^{\epsilon}
= (\sum_{i=1}^N a_i {\hat O}^{\epsilon}_{iL} + a_0) |0_r> + \ee .
\label{mdef}
\end{equation}
Although the above equation does not involve an explicit determination 
of ${\bar O}^{\epsilon}_F, \Delta O^{\epsilon}_F$, it is,
nevertheless, motivated by the requirement that 
${\bar O}^{\epsilon}_L, \Delta O^{\epsilon}_L$ be approximated by their
$SU(2)$ counterparts. To see this, note that
if we could identify the states $\Psi_0^{\epsilon}$, the above equation serves to define the 
`minimal' subspace ${\cal S}^{\epsilon *} \subset \Phi^{*}_{kin}$ which can serve as the domain of the map $\me$
i.e. \\
\noindent
${\cal S}^{\epsilon *} = \{ \Psi^{\epsilon}  \in \Phi^{*}_{kin}, \; 
\Psi^{\epsilon}=(\sum_{i=1}^N a_i {\hat O}^{\epsilon}_{iF} + a_0)\Psi_0^{\epsilon} \}$.  
Let us suppose that there exists an inner product on this space which is well approximated by the $r$-Fock inner product
$<,>$ in the following sense.
Consider $\Psi_1, \Psi_2 \in {\cal S}^{\epsilon *}$
with $\me \Psi_j = |\psi_j > + \ee, j= 1,2$. Then 
let $<,>_F$ on ${\cal S}^{\epsilon *}$ be such that
\begin{equation}
< \Psi_1, \Psi_2>_F = <\psi_1 | \psi_2 > + \ee  (\sqrt{<\psi_1|\psi_1>} +\sqrt{<\psi_2|\psi_2>} )
+ O(\epsilon^4).
\label{3.8}
\end{equation}
It is straightforward to check that for such a $<, >_F$, the equations (\ref{mphi})-(\ref{mdef}) imply that fluctuations
and expectation values of  ${\hat O}^{\epsilon}_F$ in any  state $\Psi_0^{\epsilon}$ approximate their linearised
counterparts to the desired accuracy of $\e$.

The existence of ambiguities of $\ee$  in the right hand sides of (\ref{mphi}) and (\ref{mophi})
may be motivated as follows. Equation (\ref{mophi}) may be thought of as the quantum version of
the classical relations (\ref{hH}) and (\ref{gG}). Since ${\hat O}_L^{\epsilon}|0_r>$ has norm
of $\e$, the ambiguity in the image of ${\cal M}^{\epsilon}$ must be of 
higher order than $\epsilon$ and we choose it to be $\ee$.

The rest of this work is devoted to showing the existence of at least one 
(1 parameter family of maps)
$\me$ which satisfies equations (\ref{mphi})-(\ref{mdef}).
\footnote{Strictly speaking, our results are slightly weaker in that 
  we analyse these equations subject to a mild restriction on the
loops labelling $O^{\epsilon}_L,O^{\epsilon}_F$, namely that these loops intersect 
the weave defined in section 4A in at most a finite number of points.}

\subsection{Strategy for an explicit construction of $\me$}

The approximate correspondence between linearised and full gravity motivated us to
require that the  
image of $\me$ be defined with small ambiguities of $\ee$ in equations (\ref{mphi})-(\ref{mdef}).
Similar considerations should also apply to the domain of $\me$ i.e. it seems plausible that
$\me$ should map 2 distributions  which differ by a `small' amount to the same image. Our ignorance
of any useful inner product on $\Phi^*_{kin}$ makes it difficult to define this notion of 
`smallness'. However, even though we do not know of any useful inner product we can define
 norms and semi- norms on $\Phi^*_{kin}$. Indeed, we shall define a semi- norm on $\Phi^*_{kin}$ and 
(partially) specify the `kernel' of $\me$ as containing distributions which are of $\de$ in this
semi- norm for a suitably defined small $\epsilon$ dependent parameter, $\delta_{\epsilon}$. 
We define the kernel of (the 1 parameter family of maps) $\me$  to be the set (of all 1 parameter families) 
of distributions, whose images by $\me$ are  {1 parameter families of sets of } states
with $r$- Fock norm of $\ee$.

The operators ${\hat O}_L^{\epsilon}$ satisfy the following relation in their actions on $|0_r>$
(this relation is just the `Poincar'e invariance condition' of \cite{melingrav} evaluated to $\e$):
\begin{equation}
i({\hat h}_{\alphaa} -1) |0_r>= {\hat g}_{\alphaa (r)} |0_r > + \ee .
\label{pl}
\end{equation}
This in conjunction with (\ref{mophi})  implies that 
\begin{equation}
\me (2(\sum_{k=1}^3 tr{\hat H}_{\alpha^k}\tau_k)  + {\hat G}_{\alphaa (r)})\Psi^{\epsilon}_0 = \ee .
\label{mpl}
\end{equation}

Our strategy will be to attempt to solve the following `$SU(2)$ Poincar'e invariance condition' upto ambiguities 
of $\de$
\footnote{ Actually, we solve an equation of the form 
$((2\sum_{k=1}^3tr{\hat H}_{\alpha^k}\tau_k) +{\hat G}_{\alphaa (r)} + a_{\alphaa}^{\epsilon} )\Psi^{\epsilon}_0  = \de$ where
$a_{\alphaa}^{\epsilon}$ is  complex and of $\ee$. Note that 
$a_{\alphaa}^{\epsilon}\Psi^{\epsilon}_0$ is also in the kernel of $\me$.}:
\begin{equation}
2\sum_{k=1}^3tr {\hat H}_{\alpha^k}\tau_k\Psi^{\epsilon}_0 =  - {\hat G}_{\alphaa (r)}\Psi^{\epsilon}_0  + \de ,
\label{pf}
\end{equation}
to obtain $\Psi^{\epsilon}_0$.
Once we obtain such a $\Psi^{\epsilon}_0$, we can define the action of $\me$ as follows:
\begin{eqnarray}
\me \Psi^{\epsilon}_0 &:=& |0_r> + \ee  \\
\me {\hat G}_{\alphaa (r)} \Psi^{\epsilon}_0 &:=&  {\hat g}_{\alphaa (r)} |0_r> + \ee .
\label{stratg}
\end{eqnarray}
Equation (\ref{pf}) ensures that the relevant conditions involving the connection dependent 
operators in the linearised and $SU(2)$ theories are satisfied. Finally, $\me$ can be extended by 
appropriate linearity to satisfy equation (\ref{mdef}).

\section{The weave and the semi- norm.}

The semi- norm comprises one of two auxilliary structures that we will define in order to construct 
an $\me$ satisfying equations (\ref{mphi})- (\ref{mdef}).
It turns out that in order to define the semi- norm we use, we need to define a second auxilliary structure
namely a (distributional) `weave' state. This state is the counterpart  of the flat triad $\delta^a_i$
of linearised theory. We define the weave state in section 4A and the semi- norm in section 4B.
In section 4C, we describe the general structure of the $SU(2)$ distributions which we shall consider
in the rest of this work.

\subsection{The weave state.}

The weave state is the analog of the classical flat triad $\delta^a_i$ in the 
following sense. We shall construct a set of extended spin network states
 denoted by $\{|\psi_{\Delta}>\}$ which are all based on the same 
`lattice-like' graph $\Delta$. Then the weave state, 
\footnote{ In referring to this state as a weave state, we follow the nomenclature 
used for states which correspond to classical spatial geometries \cite{weave}.
The state defined here is based on a graph identical to that used by 
Rovelli in \cite{carlo} and Zegwaard in \cite{joost}.} 
$\Psi_{weave}\in \Phi^*_{kin}$,
has the following property
\begin{equation}
({\hat E}^a_{i(r)}({\vec x})\Psi_{weave})[|\psi_{\Delta} >] = 
(1+ O^{\infty}(\epsilon ))\delta^a_i \Psi_{weave})[|\psi_{\Delta} >].
\label{Eweave}
\end{equation}
Here, $\vec x$ is restricted to be within the probe region i.e. $|{\vec x}| <Ss$ (see 3A(i) for a definition of $S$)
and $O^{\infty}(\epsilon )$ has been 
defined at the end of section 1. 
Moreover for any state $|\psi_{\Delta_{\perp}}>$ orthogonal to the span of
$\{|\psi_{\Delta}>\}$, the weave state satisfies
\begin{equation}
({\hat E}^a_{i(r)}({\vec x})\Psi_{weave})[|\psi_{\Delta_{\perp}} >] = 
\delta^a_i \Psi_{weave}[|\psi_{\Delta_{\perp}} >] =0.
\label{Eweave1}
\end{equation}
In subsection 1 below, we describe our construction of the weave state  and in 
 subsection 2, we display some of its properties. Calculations pertaining
to the explicit verification of (\ref{Eweave}) and (\ref{Eweave1}) are contained in Appendix A.

\subsubsection{Construction of the weave state}
 
Let $\Delta$ be the graph corresponding to a cubical lattice with edges along the
the cartesian coordinate directions. Let the lattice spacing be $\sqrt{\gamma l_P^2\over 2}$
and let it occupy a volume $(2L)^3$ centered about the origin. Let $L$ be such that 
$L>> s$ and  $s_{curv}>>L>>r>> l_P$.
\footnote{Note that since these inequalities are $\epsilon$ dependent, $\Delta$ and $\Psi_{weave}$
also depend on $\epsilon$. However, we shall regard this dependence as `weak' in the sense
that we hope that  a better treatment incorporating contructions appropriate to asymptotic
flatness (see for example \cite{ttasympflat}) will allow $L$ to be taken to $\infty$ without 
altering our conclusions. In this connection, we note  that the inequality 
$s_{curv}>>L $ is not crucial to the considerations of appendix A; 
as can be verified, $L>>s_{curv}$ would 
suffice equally well. In what follows, we shall refrain from explicitly denoting the
$L(\epsilon )$ dependence of $\Psi_{weave}$}.

Define the spin net $|\Delta >$ based on the graph 
$\Delta $ by:

\noindent (a) Coloring all edges of $\Delta$ with $j={1\over 2}$ i.e. with spin-half
 representations. 

\noindent (b) Defining trivial intertwiners at the vertices  of $\Delta$ i.e. mapping, trivially, 
 the 
$j={1\over 2}$ representation on an  incoming edge in the $i$th direction ($i=1,2,3$)at  any
vertex  to the $j={1\over 2}$ representation on the outgoing edge 
in the same direction (this pertains to vertices which are not at the boundary of $\Delta$).

\noindent (c) Choosing {\em any} intertwiners at the vertices on the boundary of the
lattice i.e. `tieing up' the boundary  points on the planes $x^i= \pm L$ in any 
convenient way. This choice has no bearing on our subsequent considerations.

Define the spin net $|\Delta, \{{\vec x}_k, k=1..N\}>$ as follows. Pick $N$ points 
$\{{\vec x}_k, k=1..N\}$ on the graph $\Delta$ such that they are {\em not} on the 
vertices of $\Delta$. In addition to (a), (b) and (c) above, define the intertwiners
at the new vertices $\{{\vec x}_k, k=1..N\}$ as follows. Let ${\vec x}_k$ be on an edge 
along the $i$th direction. Then the intertwiner is chosen as $2\tau_{iA}^B, \; A,B=1,2$
(clearly $\tau_{iA}^B$ maps the incoming $j={1\over 2}$ representation to the 
outgoing $j={1\over 2}$ representation at the vertex ${\vec x}_k$). Note that in terms of
the labelling of extended spin nets defined in \cite{area}, 
$\tau_{iA}^B =M_a^i C^{aB}_A$ where (i)$C^{aB}_A$ is the invariant tensor which maps
the $j={1\over 2}$ representation  to the product of the $j={1\over 2}$ and $j=1$ 
representations and (ii) $M_a^i,\; i=1,2,3$ are vectors in the $j=1$ representation 
namely, $M_a^1 =(\sqrt{3\over 8}i, 0, -\sqrt{3\over 8}i)$,
$M_a^2 =(-\sqrt{3\over 8}, 0,\sqrt{3\over 8}i)$ 
and $M_a^3 =(0,\sqrt{3\over 2}i, 0)$. 
$A,B=1,2$ are $j={1\over 2}$ indices and  $a=1,2,3$ is 
a $j=1$ index.

It can be checked that $|\Delta, \{{\vec x}_k, k=1..N\}>$ is normalised and that 
\begin{equation}
<\Delta, \{{\vec y}_j, j=1..M\}|\Delta, \{{\vec x}_k, k=1..N\}>= 0,
\label{onormal}
\end{equation}
unless $\{{\vec y}_j \}=\{{\vec x}_k\}$.  $\Psi_{weave}$ is defined as the 
distribution
\begin{equation}
\Psi_{weave}:= \sum_{N=0}^{\infty}\; \sum_{{\vec x}_k, k=1..N} (-i)^N <\Delta, \{{\vec x}_k\}|,
\label{defweave}
\end{equation}
where the second (uncountable) sum is over all possible values of $\{{\vec x}_k\}$.
Clearly, $\Psi_{weave}$ is in $\Phi^*_{kin}$ i.e. it has a well defined
action on any finite linear combination of spinnets.

\subsubsection{Properties of the weave state.}

The weave state has nice properties with respect to the action of the smeared triad operators.
The considerations of Appendix A show that equation (\ref{Eweave}) is satisfied for 
$|{\vec x}| <Ss$ (see 3A(i) for a definition of $S$).  
Although we do not display them here, straightforward calculations along the lines of Appendix A
show that for $|{\vec x}_j|<Ss, j=1..n, \;n$ independent of $\epsilon$, 
\begin{eqnarray}
\prod_{j=0}^{n-1}{\hat E}^{a_{n-j}}_{(r)i_{n-j}}({\vec x}_{n-j})  \Psi_{weave} [|\psi_{\Delta}> ]
& =& {\hat E}^{a_n}_{(r)i_n}({\vec x}_n)....
{\hat E}^{a_1}_{(r)i_1}({\vec x}_1) \Psi_{weave} [|\psi_{\Delta}> ] \nonumber \\
&=& (1 + O^{\infty}(\epsilon )) \prod_{j=0}^{n-1}\delta^{a_{n-j}}_{i_{n-j}} \Psi_{weave} [|\psi_{\Delta}> ].
\label{Enweave}
\end{eqnarray}
Here $|\psi_{\Delta}> = |\Delta >$ or $|\Delta \{{\vec x}_k\}>$.
From (\ref{Enweave}) it is straightforward to show that 
\begin{equation}
\prod_{j=0}^{n-1}
({\hat E}^{a_{n-j}}_{(r)i_{n-j}}({\vec x}_{n-j}) -\delta^{a_{n-j}}_{i_{n-j}})
 \Psi_{weave} [|\psi_{\Delta}> ] = O^{\infty}(\epsilon )\Psi_{weave} [|\psi_{\Delta}> ].
\label{hnweave}
\end{equation}
Equation (\ref{hnweave}) captures the precise sense in which the weave state is the analog of the 
flat triad $\delta^a_i$ {\em within the probe region} $|{\vec x}| < Ss$. Note however that 
since $G_{\alphaa ab(r)}$ is not, in general, a function of compact support (see Appendix B), the 
Poincare invariance condition (\ref{pl}) as well as its $SU(2)$ counterpart (\ref{pf}) involve properties
of the triad everywhere {\em including} $|{\vec x}| > Ss$. Hence to proceed further either 
(a) we need the weave state to capture properties of the flat triad  for all ${\vec x}$ or 
(b) we may attempt to restrict the Poincare invariance condition in such a way as to 
depend only on properties of the triad in the probe region.
For simplicity we shall assume that  option (a) is viable. More precisely, we assume that 
an improved treatment will yield $\Psi_{weave}$ which satisfies (\ref{hnweave}) for all
${\vec x}$ ( see also footnote 10) in this regard). While this seems to us to be a
reasonable assumption, we emphasize that its validity must be checked in a more comprehensive
treatment. We also note that preliminary calculations made by us indicate that option (b) may also 
be viable. Since the relevant analysis pertaining to option (b) is a bit involved and
may obscure our main line of argument in an already involved paper, we refrain from reporting
on this matter here.

Thus, we shall assume that (\ref{hnweave}) is valid everywhere on the spatial slice. Then, as shown
in appendix B the following identities hold for $n$ independent of $\epsilon$:
\begin{equation}
\prod_{j=0}^{n-1} {\hat G}^{\dagger}_{\alphaa_{n-j}}\Psi_{weave} [|\Delta>] = O^{\infty}(\epsilon )\Psi_{weave} [|\Delta>],
\label{gnweave1}
\end{equation}
\begin{equation}
\prod_{j=0}^{n-1} {\hat G}^{\dagger}_{\alphaa_{n-j}}\Psi_{weave} [|\Delta \{{\vec x}_k\}>] = O^{\infty}(\epsilon )\Psi_{weave} 
[|\Delta\{{\vec x}_k\}>].
\label{gnweave2}
\end{equation}

\subsection{The semi- norm.}

As discussed in \cite{melingrav}, $r$- Fock states can be thought of as $U(1)^3$ distributions
i.e. distributions  with respect to the finite span of $U(1)^3$ flux network states. This is in
close structural similarity to the LQG states under consideration,
the latter being $SU(2)$ distributions with respect to the finite span of $SU(2)$ spinnets.
Although we do not show it here, our choice of semi- norm below is motivated by 
the existence of a similar structure in the 
$U(1)^3$ theory. We shall
comment on this further in section 7.
For now, we shall simply display our choice of semi- norm and show in the subsequent section that 
our choices furnish the map ${\cal M}^{\epsilon}$ in accord with the strategy outlined in section
3C.

Consider the set of graphs $G_{\Delta}$ which intersect $\Delta$ at most in a finite number 
of points. Denote the spinnet based on the trivial graph as $|\cdot >$. Clearly, every spinnet
based on a graph $\alpha \in G_{\Delta}$ can be written in the form ${\hat {\cal N}}_{\alpha, {\vec c},{\vec M}}|\cdot >$
where ${\hat{\cal N}}_{\alpha, {\vec c},{\vec M}}$ is the generalised holonomy operator corresponding to the 
spin net labelled by the graph $\alpha$, invariant tensors ${\vec c}$ and representation vectors ${\vec M}$ \cite{area} 
(the labels defining the  colour of the edges of $\alpha$ have been suppressed as in \cite{area}). 
Explicitly, in the connection 
representation, $|\cdot >$ corresponds to the state $\psi ({\bar A})=1$ and the action of the 
 operator
${\hat {\cal N}}_{\alpha, {\vec c},{\vec M}}$ corresponds to multiplication 
by  ${\cal N}_{\alpha, {\vec c},{\vec M}}({\bar A})$ (see equation (4.19) of \cite{area}).

For any such ${\hat {\cal N}}_{\alpha, {\vec c},{\vec M}}$ consider the states 
${\hat {\cal N}}_{\alpha, {\vec c},{\vec M}}|\Delta>$ and
${\hat {\cal N}}_{\alpha, {\vec c},{\vec M}}|\Delta \{{\vec x}_k, k=1..N\}>$. Denote by $V$ the finite span of these
states for all choices of extended spinnets based on every  $\alpha \in G_{\Delta}$. 
Denote by $V_{\perp}$ the set of states which are orthogonal to
every element of  $V$. Clearly $V + V_{\perp}$ is a dense subspace of the kinematical
Hilbert space.

Next, consider the set of {\em loops} $G^s_{\Delta loop}$ 
which are of length less than or equal to $s$
\footnote{Note that unlike the loops which label $O^{\epsilon}_L,O^{\epsilon}_F$, these loops are
{\em not} restricted to lie within a volume $(Ss)^3$ about the origin (see section 3A(i)).}
 and which intersect $\Delta$ at most
in a finite number of points. For $\eta_I, \beta_J \in G^s_{\Delta loop}, \; I=1..P, \; J=1..M$,
$k_J \in \{1,2,3\}$,
define the operators
\begin{equation}
{\hat O}_{\{\eta_I,\beta_J,k_J\}} := \prod_{I=1}^P {{\rm Tr} {\hat H}_{\eta_I}\over 2} 
   \prod_{J=1}^M{\rm Tr} {\hat H}_{\beta_J}\tau_{k_J}
\;\;\;\;\;{\hat O}_{\{\eta_I\}} := \prod_{I=1}^P {{\rm Tr} {\hat H}_{\eta_I}\over 2} .
\label{defO}
\end{equation}
Define $V^s_{loop} \subset V$ to be the finite span of the states \\
\noindent
$\{ {\hat O}_{\{\eta_I,\beta_J,k_J\}} |\Delta>,{\hat O}_{\{\eta_I\}}|\Delta>,
{\hat O}_{\{\eta_I,\beta_J,k_J\}}|\Delta \{{\vec x}_k, k=1..N\}>,
{\hat O}_{\{\eta_I\}}|\Delta \{{\vec x}_k, k=1..N\}>\}$ for all choices of
${\{\eta_I,\beta_J,k_J,{\vec x}_k\}}$ and finite $P,M,N$.

For any $\Psi \in \Phi^*_{kin}$ we define the semi- norm
\begin{equation}
||\Psi||^{\epsilon} = \sup_{| \psi>}|\Psi [|\psi >]|
\label{defseminorm}
\end{equation}
where $| \psi>$ ranges over the following:\\
\noindent (a)$| \psi>$ is a spinnet in  $V_{\perp}$, 

\noindent (b)$| \psi \rangle$ is $ (M+P)^{-(M+P)}{\hat O}_{\{\eta_I,\beta_J,k_J\}} |\Delta \rangle$ or 
$P^{-P}{\hat O}_{\{\eta_I\}}|\Delta \rangle$
or $(M+P)^{-(M+P)}{\hat O}_{\{\eta_I,\beta_J,k_J\}}|\Delta \{{\vec x}_k, k=1..N\}>$ or 
$P^{-P}{\hat O}_{\{\eta_I\}}|\Delta \{{\vec x}_k, k=1..N\}>$,  for all choices of
${\{\eta_I,\beta_J,k_J, {\vec x}_k\}}$ and finite $P,M,N$.

$||\;\;||^{\epsilon}$ is a semi- norm precisely because $|\psi >$ is not allowed to range over the 
orthogonal complement of $V^s_{loop}$ in $V$.
The above semi- norm is 
$\epsilon$- dependent both due to the  (weak) dependence of $\Delta$ on $\epsilon$ 
(see footnote 10) as well as the dependence of $s>>r$ on $\epsilon$.
The $\epsilon$- dependence of the semi-norm is denoted by the superscript $\epsilon$.

\subsection{General structure of the LQG distributions under study.}

Consider an arbitrary distribution $\Phi \in \Phi^*_{kin}$. Define $\Psi \in\Phi^*_{kin}$
from $\Phi, \Psi_{weave}$ as follows.\\
\noindent (a) Let $|\psi > \in V_{\perp}$ be a spinnet. Then $\Psi [ \psi ] :=0$. \\
\noindent (b) Let $\alpha \in G_{\Delta}$ and consider any `spinnet' operator
${\hat {\cal N}}_{\alpha, {\vec c},{\vec M}}$ as defined in section 4.2 above. Then
$\Psi [{\hat {\cal N}}_{\alpha, {\vec c},{\vec M}} |\Delta > := \Psi_{weave} [|\Delta >] 
\Phi [{\hat {\cal N}}_{\alpha, {\vec c},{\vec M}} |\cdot >]$
and  $\Psi [{\hat {\cal N}}_{\alpha, {\vec c},{\vec M}} |\Delta\{{\vec x}_k\} > 
:= \Psi_{weave} [|\Delta \{{\vec x}_k\}> ] \Phi [{\hat {\cal N}}_{\alpha, {\vec c},{\vec M}} |\cdot >]$.
This defines the action of $\Psi $ on a basis of $V$. \\
\noindent (c) Extend the action of $\Psi$  as defined in (a) and (b) to the finite span
of all spinnets by linearity.
The $SU(2)$ states of interest in the rest of this work will have the structure of  $\Psi$ defined above.

\section{The vacuum as an $SU(2)$ distribution.}

Below, we exhibit solutions to the $SU(2)$ Poincare invariance condition (\ref{pf}) (modulo
terms in the kernel of $\me$ (see footnote 8, section 3C in this regard)),
thus implementing the strategy of section 3C. The particular choice of our solutions 
as well as that of $\delta_{\epsilon}$ is motivated  by their analogs in the description
of $r$- Fock states by $U(1)^3$ distributions. We shall comment on this further in section 7.
Here we shall simply display our choices and show that they furnish the required $\me$.


\subsection{Explicit form of the solution to the Poincare invariance condition.}

The putative solution $\Psi_0^{\epsilon}$ is of the type described in section 4C with 
$\Phi = \Phi_0^{\epsilon}$ where $\Phi_0^{\epsilon}$ is defined as
\begin{eqnarray}
\Phi_0^{\epsilon} &=& \delta_{A=0} + {\hat E^2}\delta_{A=0}, \nonumber\\
{\hat E^2}&:=&
{i\over 2\gamma l_P^2}\int d^3k k 
[(1+2i\gamma ) ({\hat E}_{(r)}^{+}({\vec k}))^{\dagger}{\hat E}_{(r)}^{+}({\vec k})
-(1-2i\gamma)({\hat E}_{(r)}^{-}({\vec k}))^{\dagger}{\hat E}_{(r)}^{-}({\vec k})] ,
\label{defsu2vac}
\end{eqnarray}
where $\delta_{A=0}$ is  the distribution corresponding to the connection $A_a^i=0$ i.e.
for any $\phi ({\bar A}) \in {\cal L}^2 ({\bar {\cal A}}, d\mu_0)$,
\begin{equation}  
\delta_{A=0}[\phi ({\bar A})] := \phi ({\bar A})|_{{\bar A}=0}.
\label{defa0}
\end{equation}

Consider a set of operators ${\hat O}_i, i= 0..n$ such
that each ${\hat O}_i$ is of the type defined in equation (\ref{defO}).
Then we have the following key identity which is proved in Appendix D.
\begin{equation}
{\hat O}_0 {\hat G}_{\alphaa_1(r)}{\hat O}_1...{\hat G}_{\alphaa_n(r)}{\hat O}_n\Psi^{\epsilon}_0
=\Psi^{\epsilon}  + O^{\infty}(\epsilon ).
\label{oeoe1}
\end{equation}
Here $\Psi^{\epsilon}$ is a distribution of the type defined in section 4C with
$\Phi =\Phi^{\epsilon}$  where $\Phi^{\epsilon}$ is defined as  
$\Phi^{\epsilon}  := {\hat O}_0(\int G_{\alphaa_1(r)a_1b_1}{\hat E}^{a_1b_1}_{(r)}){\hat O}_1
..(\int G_{\alphaa_n(r)a_nb_n}{\hat E}^{a_nb_n}_{(r)}){\hat O}_n\Phi^{\epsilon}_0$. 

The above identity (\ref{oeoe1}) states that distribution on its left hand side is the sum of 
$\Psi^{\epsilon}$ and a distribution of semi-norm $O^{\infty}(\epsilon)$. 
Since $\Psi^{\epsilon}$ is also of the type defined in section 4C, this identity allows us to
rewrite the Poincare invariance condition as a condition on
the `$\Phi$' part of the distribution. In this regard, it is useful
to define the following seminorm, $||\;||^{\epsilon}_1$, pertinent to $\Phi$:
\begin{equation}
||\Phi ||^{\epsilon}_1 = \sup_{|\phi >} |\Phi [|\phi >]|,
\label{seminorm1}
\end{equation}
where $|\phi> \in \{(M+P)^{-(M+P)}{\hat O}_{\{\eta_I,\beta_J,k_J\}}|\cdot >,P^{-P}{\hat O}_{\{\eta_I\}}|\cdot >\}$
for all choices of ${\{\eta_I,\beta_J,k_J\}}$ and finite $P,M,N$ (${\hat O}_{\{\eta_I,\beta_J,k_J\}},{\hat O}_{\{\eta_I\}}$
are defined by (\ref{defO})). Note that for any distribution $\Psi$ of the type defined in section 4C, we have that 
\begin{equation}
||\Psi ||^{\epsilon} = || \Phi ||^{\epsilon}_1 .
\label{equivsemi}
\end{equation}
Then it follows from (\ref{oeoe1}) that the $SU(2)$ Poincare invariance condition (\ref{pf})
on $\Psi^{\epsilon}_0$, {\em for loops $\alpha^k \in G^s_{\Delta loop}$} (and which are, as in (\ref{pf}) confined to
a volume $(Ss)^3$ about the origin),
 is equivalent to the following condition on $\Phi^{\epsilon}_0$:
\begin{equation}
2\sum_{k=1}^3tr {\hat H}_{\alpha^k}\tau_k\Phi^{\epsilon}_0 =  
- (\int d^3x G_{\alphaa (r)ab}({\vec x}){\hat E}^{ab}_{(r)}({\vec x}))\Phi^{\epsilon}_0  + \de .
\label{pf1}
\end{equation}
Here the last term denotes a distribution with seminorm $||\;||^{\epsilon}_1$ of $\de$.
In the next subsection we show that $\Phi^{\epsilon}_0$ does indeed satisfy 
(\ref{pf1}) upto a term of $\ee$ which is in the kernel of $\me$. More precisely,
we show that 
\begin{equation}
2\sum_{k=1}^3tr {\hat H}_{\alpha^k}\tau_k\Phi^{\epsilon}_0 =  
- (\int d^3x G_{\alphaa (r)ab}({\vec x}){\hat E}^{ab}_{(r)}({\vec x}))\Phi^{\epsilon}_0  + 
\epsilon^2 c_{\alphaa}\Phi_0^{\epsilon} +\de ,
\label{pf11}
\end{equation}
where $c_{\alphaa}$ is a complex number of $O(1)$ depending only on $\alphaa$. Since 
$\me  \epsilon^2 c_{\alphaa}\Phi_0^{\epsilon} = \epsilon^2 c_{\alphaa}|0_r>$, this extra term is in
the kernel of $\me$.
This is equivalent to showing that $\Psi^{\epsilon}_0$
satisfies (\ref{mpl}) for $tr{\hat H}_{\alpha^k}\tau_k, {\hat G}_{\alphaa (r)}$ labelled by loops
$\alpha^k$ each of which are of length at most equal to $s$, are confined to 
a volume $(Ss)^3$ about the origin, {\em and} which intersect $\Delta$ in at 
most a finite number of points. Note that this is a slightly weaker condition than (\ref{mpl}) which
was defined for all loops with length at most $s$ confined to
a volume $(Ss)^3$ about the origin {\em irrespective of their 
intersections with $\Delta$}. We shall only impose this slightly weaker condition of 
Poincare invariance on $\Psi^{\epsilon}_0$ rather than the stronger one (\ref{mpl}).

\subsection{Verification of (\ref{pf11}).}

We set $\delta_{\epsilon}= \epsilon^4$. We shall explain this choice later in this subsection.
From (\ref{ehatpsi1}) it follows that 
\begin{equation}
{\hat E}^a_{(r)i}({\vec k}) (2\sum_{k=1}^3
tr (H_{\alpha^k}(A)\tau^k))|_{A=0} = -il_P^2\gamma X^{a}_{(r)\alpha^i}({\vec k}).
\label{eHX}
\end{equation}
Also, setting $n=0,N=1$ and choosing $O_1^H=tr (H_{\alpha^k}(A)\tau^k))$  in (\ref{5}) it follows that 
\begin{equation}
{\hat E^2}\sum_{k=1}^3tr (H_{\alpha^k}(A)\tau^k)|_{A=0} = \epsilon^2 c_{\alphaa}, 
\label{e2H}
\end{equation}
where $c_{\alphaa}$ is a constant of $O(1)$ which depends only on the triplet of loops $\alphaa$.

The action of the left hand side of (\ref{pf11}) on $|\phi > $ of the type appearing in (\ref{seminorm1}) is
\begin{equation}
L.H.S. := \sum_{k=1}^3tr (H_{\alpha^k}(A)\tau^k)\phi (A)|_{A=0} 
+{\hat E^2}\sum_{k=1}^3tr (H_{\alpha^k}(A)\tau^k)\phi (A)|_{A=0} 
\label{lhs1}
\end{equation}
where we have denoted the ket $|\phi >$ by the corresponding wave function in the connection representation,
$\phi (A)$. Since $tr (H_{\alpha^k}(A)\tau^k)$ vanishes at $A=0$, the first term in (\ref{lhs1}) vanishes
and in the second term either both triad operators of ${\hat E^2}$ act on $tr (H_{\alpha^k}(A)\tau^k)$ or one acts
on $tr (H_{\alpha^k}(A)\tau^k)$ and one on $\phi (A)$. Using (\ref{eHX}) and (\ref{e2H}) to evaluate these
contributions, we obtain
\begin{eqnarray}
L.H.S. &=& - (\int G_{\alphaa (r)ab}{\hat E}^{ab}_{(r)})^{\dagger}\phi (A)|_{A=0}
+\epsilon^2c_{\alphaa}\phi (A)|_{A=0} \nonumber\\
&=& - (\int G_{\alphaa (r)ab}{\hat E}^{ab}_{(r)})^{\dagger}\phi (A)|_{A=0}
+\epsilon^2c_{\alphaa}(\Phi_0^{\epsilon}[\phi] - {\hat E^2}\delta_{A=0}[\phi])\nonumber\\
&=& - (\int G_{\alphaa (r)ab}{\hat E}^{ab}_{(r)})^{\dagger}\phi (A)|_{A=0}
+\epsilon^2c_{\alphaa}\Phi_0^{\epsilon}[\phi] + O(\epsilon^4),
\label{lhs2}
\end{eqnarray}
where we have used (\ref{5}) to obtain the $O(\epsilon^4)$ term. Using  
(\ref{5}) again, the action of the right hand side of (\ref{pf11}) on $\phi $ is
\begin{equation}
R.H.S. =  - (\int G_{\alphaa (r)ab}{\hat E}^{ab}_{(r)})^{\dagger}\phi (A)|_{A=0} + O(\epsilon^4).
\label{rhs}
\end{equation}
This completes the verification of (\ref{pf11}).

Note that from (\ref{4}) and (\ref{5}) it is easy to see that 
$(\int d^3x G_{\alphaa (r)ab}({\vec x}){\hat E}^{ab}_{(r)}({\vec x}))\Phi^{\epsilon}_0$ is $\ee$.
But the image of this quantity by $\me$ is expected to be (see equation(\ref{stratg})) 
${\hat g}_{\alphaa (r)}|0_r>$. Since the norm of the latter is expected to be of $\e$ from 
(\ref{2.19}), we do not want $(\int d^3x G_{\alphaa (r)ab}({\vec x}){\hat E}^{ab}_{(r)}({\vec x}))\Phi^{\epsilon}_0$
to be in the kernel of $\me$. Hence we choose $\delta_{\epsilon}<< \epsilon^2$ and our choice is 
$\delta_{\epsilon} = \epsilon^4$.

This argument is merely a plausability one. Strictly speaking, we need to show that for every $\alpha^k$ such that 
$\Delta g_{\alphaa (r)} \approx \epsilon$ to nontrivial leading order in $\epsilon$, there exists
$|\phi >$ of the type in (\ref{seminorm1}) such that 
$(\int d^3x G_{\alphaa (r)ab}({\vec x}){\hat E}^{ab}_{(r)}({\vec x})) \Phi^{\epsilon}_0[|\phi >]\approx \epsilon^2$ 
to nontrivial leading order in $\epsilon$. It is straightforward to see that this is indeed true for
$|\phi > = \sum_{k=1}^3 tr( {\hat H}_{\alpha^k}\tau^k) |\cdot >$. This suggests that we generalise the definition 
of the seminorm so as to allow $|\phi >$ of this type. While we do not anticipate any obstruction to such a trivial 
generalisation, we leave a thorough verification of this to future work.

\subsection{Higher order operators.}

It can be verified that the action of quadratic and higher order products of operators of type 
${\hat O}_L^{\epsilon}$ on $|0_r>$ yields states of norm at most $\ee$ (this is under the assumption that the 
number of operators of type ${\hat O}_L^{\epsilon}$ in the product is independent of $\epsilon$).
Therefore, it is natural to demand that the action of their $SU(2)$ counterparts map the $SU(2)$ vacuum 
into the kernel of $\me$. Also note that the $SU(2)$ function, $(Tr (H_{\alpha})-1)$, $\alpha\in G^{s}_{\Delta loop}$
 has leading order term
quadratic in the connection. Hence, it is natural to also demand that 
$((Tr {\hat H}_{\alpha})-1)\Psi^{\epsilon}_0$
be in the kernel of $\me$.

Indeed, using the key identities (\ref{oeoe1}), (\ref{4}) and (\ref{5}), it is straightforward to verify that 

\noindent (i) $((Tr {\hat H}_{\alpha})-1)\Psi^{\epsilon}_0 = \epsilon^2 d_{\alpha}\Psi^{\epsilon}_0 + \de$,
where $d_{\alpha}$ is an $\alpha$- dependent complex number of $O(1)$ and $\alpha \in G^{s}_{\Delta loop}$.

\noindent (ii) The action of any product of 2 operators each of which belongs to the set \\
\noindent $\{ O^{\epsilon}_F,((Tr {\hat H}_{\alpha})-1),\; \alpha \in G^{s}_{\Delta loop}\}$, on 
$\Psi^{\epsilon}_0$, yields distributions in the kernel of $\me$. Specifically, these distributions are
either of type $\ee \Psi^{\epsilon}_0$ or of $\de$.

\noindent (iii) The action of any product of $n$ operators ($n>2$, $n$ independent of $\epsilon$) each of which is in 
the set defined in (ii) above, on $\Psi^{\epsilon}_0$, yields a distribution of $\de$.

\section{The linearised constraints in the $SU(2)$ theory.}

We present qualitative arguments to show that the $SU(2)$ vacuum state
is mapped to  a distribution of $\de$ by the linearised constraints (\ref{lgauss})-(\ref{lscalar})
expressed as operators in the $SU(2)$ theory. 

We first discuss the purely connection dependent
linearised vector and scalar constraints $V^L$ and $C^L$. We shall impose these constraints everywhere on the 
spatial slice except on the set of measure zero containing points on the weave $\Delta$.
This implies that in the $SU(2)$ theory it should be possible to regularise ${\hat V}^L,{\hat C}^L$ in terms
of holonomies along edges which do not intersect $\Delta$. From the structure of $\Psi_0^{\epsilon}$ 
defined in section 4C, this implies that the action of the regularised operators (we denote these by 
${\hat V}^L_{reg},{\hat C}^L_{reg}$) on $\Psi_0^{\epsilon}$ is simply to change 
$\Phi_0^{\epsilon}$  to ${\hat V}^L_{reg}\Phi_0^{\epsilon}$, ${\hat C}^L_{reg}\Phi_0^{\epsilon}$.
Note also that classically, the positive and negative helicity parts of the triad Poisson commute with 
$V^L,C^L$. Hence in any reasonable regularization, we expect that in the limit of the regulator
being removed, we have that 
\begin{equation}
[{\hat V}^L, {\hat E^2}]=[{\hat V}^L, {\hat E^2}]=0 .
\label{ve2ce2}
\end{equation}
Finally, since $V^L,C^L$ vanish when evaluated at the zero connection, we expect that 
\begin{equation}
{\hat V}^L\delta_{A=0} = {\hat C}^L\delta_{A=0} = 0.
\label{vdcd}
\end{equation}
Equations (\ref{ve2ce2}) and (\ref{vdcd}) imply that ${\hat V}^L,{\hat C}^L$ annihilate 
$\Phi^{\epsilon}_0$, and in view of the above discussion also annihilate $\Psi^{\epsilon}_0$.

Next, we examine the linearised Gauss Law constraint $G_i^L$ (see (\ref{lgauss}).
Since in our work the basic triad operator is ${\hat E}^a_{(r)i}$, we define the Gaussian smeared
constraint $G^L_{(r)i}$ by 
\begin{eqnarray}
G^L_{(r)i} ({\vec x}) &=& \int d^3y  G_i^L
({\vec y}){e^{-{|{\vec x}- {\vec y}|^2 \over 2 r^2}}\over (2\pi r^2)^{3\over 2}}
\label{glr1}
\\
&=& \partial_a e^a_{(r)i}({\vec x}) + \epsilon_i^{\;ja}A_{(r)aj}({\vec x}).
\label{glr2}
\end{eqnarray}
The first term in (\ref{glr2}) is obtained after a by parts integration and $A_{(r)aj}({\vec x})$ is the 
Gaussian smeared connection. 
In the $SU(2)$ theory our arguments pertaining to $V^L,C^L$ apply to the connection dependent part of 
(\ref{glr2}). Hence we expect that for any reasonable definition of the connection part of 
$G^L_{(r)i}({\vec x})$, this connection part annihilates $\Psi^{\epsilon}_0$. Hence we need analyse only 
the triad dependent part of (\ref{glr2}). 

At the classical level we would like to impose the linearised constraints as $\e$ restrictions on the 
phase space.  Since (\ref{glr1}) has the dimensions of inverse length, we need to integrate it against 
a gauge parameter $\Lambda^i({\vec x})$ which has dimensions of inverse length square so as to obtain
the dimensionless quantity $G^L_{(r)}(\Lambda ):= \int d^3x \Lambda^iG^L_{(r)i}$. We denote the triad 
dependent part of $G^L_{(r)}(\Lambda )$ by 
\begin{equation}
G^L_{(r)e}(\Lambda ):= -\int d^3x (\partial_a \Lambda^i({\vec x}) )  e^a_{(r)i}({\vec x}),
\end{equation}
where we have performed a by parts integration and assumed that $\Lambda^i({\vec x})$ is of compact support.
We shall meet the requirement $|G^L_{(r)}(\Lambda )| \sim \e$ by imposing the (sufficient) condition
\begin{equation}
\int d^3x |\partial_a \Lambda^i({\vec x})| \sim O(1).
\label{lambda}
\end{equation}

In the $SU(2)$ theory we have 
\begin{equation}
{\hat G}^L_{(r)e}(\Lambda )\Psi^{\epsilon}_0 =-\int d^3x (\partial_a \Lambda^i({\vec x}) )  
({\hat E}^a_{(r)i}({\vec x})- \delta^a_i)\Psi^{\epsilon}_0 .
\end{equation}
Using (\ref{Eweave}) and (\ref{lambda}), we obtain
\begin{equation}
|{\hat G}^L_{(r)e}(\Lambda )\Psi_{weave}[|\psi_{\Delta}>]| = O^{\infty}(\epsilon ),
\label{glweave}
\end{equation}
where $|\psi_{\Delta}>] \in \{|\Delta>, |\Delta \{{\vec x}_k\}>\}$.
From equation (\ref{glweave}) and the structure of $\Psi^{\epsilon}_0$ as defined in 
section 4C, we have that 
\begin{equation}
{\hat G}^L_{(r)e}(\Lambda )\Psi^{\epsilon}_0 =\de \Leftrightarrow {\hat G}^L_{(r)E}(\Lambda )\Phi^{\epsilon}_0= \de
\label{psiphi}
\end{equation}
where ${\hat G}^L_{(r)E}(\Lambda ):= -\int d^3x (\partial_a \Lambda^i({\vec x}) )  
{\hat E}^a_{(r)i}({\vec x})$ and 
where $\de$ in the first equality is defined with respect to $||\;||^{\epsilon}$ and in the second equality
with respect to $||\;||^{\epsilon}_1$ (see (\ref{seminorm1})).

Using the (inverse Fourier transform of) equation (\ref{ehatpsi1}) it is straightforward to obtain
\begin{equation}
{\hat G}^L_{(r)E}(\Lambda ) tr (H_{\alpha}(A)\tau^i)|_{A=0} = -\int (\partial_a\Lambda^i)X^a_{(r)\alpha}
                                  = \int (\partial_a X^a_{(r)\alpha})\Lambda^i =0,
\label{delxa1}
\end{equation}
where we have used the fact that the smeared loop form factor is divergence free i.e.
\begin{equation}
\partial_a X^a_{(r)\alpha} = 0.
\label{delxa}
\end{equation}
Moreover, equation (\ref{ehatpsi1}) also implies that 
\begin{equation}
{\hat G}^L_{(r)E}(\Lambda ) tr (H_{\alpha}(A))|_{A=0}= 0.
\label{gtrh}
\end{equation}
Equations (\ref{delxa1}) and (\ref{gtrh}) imply that 
\begin{equation}
\delta_{A=0} [{\hat G}^L_{(r)E}(\Lambda )|\phi >]=0
\label{gdelta}
\end{equation}
where $|\phi >$ is of the type defined below (\ref{seminorm1}).

Using methods similar to those in  Appendix C, it is straightforward to verify that for 
$O^H_{\alpha}\in \{tr (H_{\alpha}\tau^i), {tr H_{\alpha}\over 2}, \alpha \in G^s_{\Delta loop}\}$ we have that,
\begin{equation}
|{\hat E^2}{\hat G}^L_{(r)E}(\Lambda )O^H_{\alpha}(A)|_{A=0}< \epsilon^2 d_0 l_P^2\gamma 
               \int d^3x |\partial_a \Lambda_i ({\vec x})||X^a_{(r)\alpha}({\vec x})|,
\end{equation}
where $d_0$ is defined in (\ref{5}). Using (\ref{lambda}) and the bound \\
$|X^a_{(r)\alpha}({\vec x})|=|\oint_{\alpha}ds {\dot \alpha}^a
{e^{-{|{\vec x}- {\vec \alpha (s)}|^2 \over 2 r^2}}\over (2\pi r^2)^{3\over 2}}|< {s\over (2\pi r^2)^{3\over 2}} $,
we obtain
\begin{equation}
|{\hat E^2}{\hat G}^L_{(r)E}(\Lambda )O^H_{\alpha}(A)|_{A=0} < \epsilon^2d_0{sl_P^2\gamma \over r^3} = O(\epsilon^4) {r\over s}
\label{rbys}
\end{equation}
From the above equation and the fact that $s>>r$ it follows straightforwardly that 
\begin{equation}
|{\hat E^2}{\hat G}^L_{(r)E}(\Lambda )\prod_{I=1}^NO^H_I(A)|_{A=0} < N^3 O(\epsilon^4).
\label{x}
\end{equation}
Finally, it is easily verified that the second equality in equation (\ref{psiphi}) follows from (\ref{gdelta}) and (\ref{x}),
thus implying that the first equality in (\ref{psiphi}) holds.
This completes our arguments in support of the hypothesis that the linearised constraints expressed as 
operators in the $SU(2)$ theory map the $SU(2)$ vacuum to the kernel of $\me$.

\section{Generalization to a norm?}

In this section, we enquire as to whether our considerations admit a generalization to a norm. We find that the 
linearised Gauss constraint provides the only obstruction to 
the most obvious such generalization. In order to obtain insight
into why this happens, we are led to a comparision between the $SU(2)$ structures hitherto defined and the 
$U(1)^3$ structures (alluded to in section 1) which provided the motivation for these definitions.

$||\;||^{\epsilon}$ is not a norm because the states $|\psi >$ defined in (a) and (b) of section 4B do not
span the kinematic Hilbert space. This can be remedied by requiring that the loops $\eta_I, \beta_J$  in
section 4B (a), (b) be replaced by {\em edges}. More precisely, define $G^s_{\Delta edge}$ such that every 
element of $G^s_{\Delta edge}$ is an analytic edge of length at most $s$ which  intersects $\Delta$ in at
most a finite number of points. Since higher spin representations can be built up as products of spin ${1\over 2}$
representations and since edge holonomies along longer edges are products of those along shorter edges, it follows
that $|\psi >$  in section 4B (a), (b) defined  for $\eta_I , \beta_J \in G^s_{\Delta edge}$ span the kinematic 
Hilbert space. Thus, with this modification $||\;||^{\epsilon}$ becomes  a norm. Let us denote this norm
by $||\;||_{norm}^{\epsilon}$.  Remarkably, all our results with the single exception of the treatment 
of $G^L_i$ go through in this norm. Specifically, (\ref{delxa}) no longer holds if $\alpha$ is an edge and not
a loop. 

As mentioned in section 1, the reason our $SU(2)$ constructions work is due to their structural analogy with 
the corresponding $U(1)^3$ ones in the description  \cite{melingrav} of $r$- Fock states as $U(1)^3$ distributions.
Hence, it is instructive to view the above failure to impose the linearised Gauss law as an approximate
constraint in the $SU(2)$ theory in the context of this structural analogy. Therefore we briefly 
decribe the relevant $U(1)^3$ structures below. 

\noindent (1) From \cite{melingrav}, $|0_r>$ may be written as the $U(1)^3$ distribution $\Phi_0^L$ defined by
\begin{equation}
\Phi^L_0 := \sum_{\alphaa, \bf{\{q\}}} c_{0\alphaa, \bf{\{q\}}}
                     <\alphaa, \bf{\{q\}}|,
\label{7.1}
\end{equation}
\begin{equation}
c_{0\alphaa, \bf{\{q\}}}
= \exp\big({-i{\gamma\over 2}\int d^3x 
G_{\alphaa, \bf{\{q\}}(r)ab}({\vec x})})^*
X^{ab}_{\alphaa, \bf{\{q\}}(r)}({\vec x})
\big),
\label{7.2}
\end{equation}
where $|\alphaa, \bf{\{q\}}>$ is a $U(1)^3$  gauge invariant flux network state based on the
triplet of graphs $\alphaa$ labelled by the integers $\bf{\{q\}}$ and $G_{\alphaa, \bf{\{q\}}(r)ab}$ is essentially
the same as $G_{\alphaa, (r)ab}$ defined by (\ref{gab}) in this paper. See \cite{melingrav} for the precise
definitions of $G_{\alphaa, \bf{\{q\}}(r)ab},X^{ab}_{\alphaa, \bf{\{q\}}(r)}$. Note that $\gamma_0$ in 
\cite{melingrav} is related to $\gamma$ of this work by $\gamma_0= 2\gamma$.

\noindent (2) Consider the algebraic dual $\Phi^{*L}_{kin}$ to the finite span of extended  (i.e. not necessarily
gauge invariant)  $U(1)^3$ flux nets. For any $\Phi \in \Phi^{*L}_{kin}$ define the seminorm
$||\Phi ||_{seminorm}^L = \sup_{|\phi >} |\Phi [|\phi >]|$ with $|\phi >:= {\hat h}_{\alphaa} |\cdot >_L$, the length of 
each $\alpha^i$ being at most $s$, and with  $|\cdot >_L$ denoting the flux network state corresponding to 
the trivial loop. Then the right hand side of (\ref{7.1}) can be expanded in this semi- norm as 
\begin{eqnarray}
\Phi^L_0 &=& \Phi^L_{0approx}  + O( \epsilon^4) \label{7.3} \\
\Phi^L_{0approx}&: = &\delta_{A=0} + {\hat e^2}\delta_{A=0}, \label{7.3b}\\
{\hat e^2}&:=&
{i\over 2\gamma l_P^2}\int d^3k k 
[(1+2i\gamma ) ({\hat e}_{(r)}^{+}({\vec k}))^{\dagger}{\hat e}_{(r)}^{+}({\vec k})
-(1-2i\gamma)({\hat e}_{(r)}^{-}({\vec k}))^{\dagger}{\hat e}_{(r)}^{-}({\vec k})] ,
\label{7.3c}
\end{eqnarray}
where $\delta_{A=0}$ is  the $U(1)^3$ distribution corresponding to the $U(1)^3$ connection $A_a^i=0$.

\noindent (3) For any $\Phi \in \Phi^{*L}_{kin}$ , define the (stronger than $||\;||^L_{seminorm}$)  seminorm 
\begin{equation}
||\Phi ||^L := \sup_{|\phi >} |\Phi [|\phi >]|
\label{7.4}
\end{equation}
 with $|\phi >:= \prod_{I=1}^N{{\hat h}_{\alphaa_I} |\cdot >_L\over N^{3N}}$ for $\alpha^k_I, k=1..3,I=1..N$ being 
loops of length at most $s$, $N$ finite. Then $\Phi^L_{0approx}$ satisfies the Poincare invariance condition
(\ref{pl}) modulo distributions of $O(\epsilon^4) $ in the seminorm (\ref{7.4}) and modulo terms
of type $\ee \Phi^L_{0approx}$.

From (1)- (3) it is clear as to how $U(1)^3$ structures motivated our definitions of $SU(2)$ structures.
In particular the $U(1)^3$ structures ${\hat e}^a_i = -il_P^2\gamma {\delta \over \delta A_a^i},
\Phi_0^L$ and $||\;||^L$ bear a close analogy to the $SU(2)$ structures 
${\hat E}^a_i = -il_P^2\gamma {\delta \over \delta A_a^i},
\Phi_0^{\epsilon}$ and $||\;||^{\epsilon}$. In this context the failure of 
$||{\hat G}^L_{(r)}(\Lambda ) \Psi^{\epsilon}_0||^{\epsilon}$ to be of $\de$ may be traced, as indicated
below, to the difference between the gauge groups $U(1)^3$ and $SU(2)$.

Consider the algebraic dual $\Phi^{*L\;inv}_{kin}$ to the finite span of gauge invariant 
$U(1)^3$ fluxnets. Since larger loop $U(1)^3$ holonomies 
can be constructed as products of smaller loop ones, it follows that $|\phi >$ of the type 
in (\ref{7.4}) span the space of $U(1)^3$ gauge invariant fluxnets. Hence, although $||\;||^L$
is a seminorm on $\Phi^{*L}_{kin}$, it  is a {\em norm} on  $\Phi^{*L\;inv}_{kin}$.
Moreover, since ${\hat g_{\alphaa (r)}}, {\hat h}_{\alphaa}$ are $U(1)^3$ gauge invariant, the 
entire treatment of the $U(1)^3$ theory can be done at the gauge invariant level, and hence, in the 
context of a norm.

In contrast the very notion of linearization of the full $SU(2)$ theory as implemented in 
\cite{ars,melingrav} involves an $SU(2)$ gauge {\em variant} background. Since the 
$SU(2)$ counterparts of ${\hat g_{\alphaa (r)}}, {\hat h}_{\alphaa}$ are themselves 
{\em not} $SU(2)$ gauge invariant, it seems appropriate to use gauge variant kinematical 
states $|\psi >$ in the definition of $||\;||^{\epsilon}$. Ultimately, however, 
only $SU(2)$ {\em invariant} distributions are physically relevant in LQG. Hence one may further
restrict
attention in (\ref{seminorm1})  to only those $|\phi >$ 
which are gauge invariant in the hope that the resulting seminorm on $\Phi^*_{kin}$
reduces to a norm on $\Phi^{*inv}_{kin}$, where $\Phi^{*inv}_{kin}$ is the algebraic dual 
to the finite span of spin nets which are gauge invariant with respect to 
$SU(2)$ transformations within the probe region. However, this
hope is not realised because, in contrast to the $U(1)^3$ case, due to the non- abelian
nature of $SU(2)$, traces of large loop holonomies are not expressible as (sums of) products of 
traces of small loop holonomies. Moreover, with this restriction, it seems that 
$||{\hat G}_{\alphaa (r)} \Psi^{\epsilon}_0||^{\epsilon} = O(\epsilon^4 )$ which implies
that a change in our choice of $\delta_{\epsilon}$ to $\delta_{\epsilon}<< \epsilon^4$
would also be needed.

An alternate strategy would be to, in analogy with
$U(1)^3$ theory, restrict attention to only such
$|\psi >$ in (\ref{defseminorm}) which satisfy ${\hat G}^L_{(r)}(\Lambda )|\psi > =0$
i.e. try to impose the $U(1)^3$ constraint as an $SU(2)$ operator constraint.
Our intuition is that no solutions to such an equation exist because $SU(2)$ spinnets
are ``attuned'' to $SU(2)$ gauge transformations rather than $U(1)^3$ ones. Instead, since
in the linearised theory we are interested in  connections close to $A_a^i=0$, 
we may attempt to solve
${\hat G}^L_{(r)E}(\Lambda )\phi(A)|_{A=0}= 0$. Indeed, viewed in this way, 
$|\phi >$ defining $||\;||^{\epsilon}_1$ are precisely of this type. 

On account of the above discussion, our view is that \\
\noindent (a) the failure of (\ref{delxa})
in the context of $||\;||^{\epsilon}_{norm}$ is not an accident; rather it is indicative 
of the difference between the gauge groups $U(1)^3$ and $SU(2)$

\noindent (b)the seminorm $||\;||^{\epsilon}$ seems to be  a reasonable structure to use.
\footnote{It may be verified that 
$||G^L_{(r)E}(\Lambda ) \Psi_0^{\epsilon}||^{\epsilon}_{norm} = O(\epsilon^{2}){r\over s}$. 
This suggests the possibility of 
 exploiting the freedom in (b) of Appendix A to either redefine 
$\delta_{\epsilon}= \epsilon^2 {r\over s}$ for ${r\over s} >\epsilon^2$ or to choose 
${r\over s}\leq \epsilon^2$.  
However, since the $O(\epsilon^{2}){r\over s}$ bound depends crucially on (in our view the unduly
weak premise) equation (\ref{lambda}), we believe that such a strategy may be inappropriate.}

Neverthless, a (putative) perturbative treatment of the constraints would be more 
powerful in the context of a stronger seminorm. In view of the discussion above it would be interesting to see
if we could enlarge the definition of $|\phi>$ in (\ref{seminorm1}) such that 
(i) ${\hat G}^L_{(r)E}(\Lambda )\phi(A)|_{A=0}= 0$ (ii)  $|\phi >$ 
 although not themselves necessarily gauge invariant, do span
 the space of $SU(2)$ gauge invariant spinnets based on graphs in $G_{\Delta}$.

\section{Discussion}

One of the feautures of our work is the, at first perhaps surprising but nevertheless essential, role of the 
probe scale $s$. At the classical level $s$ acts as an infrared cut off, allowing  us to ignore large distance effects 
leading to the formation of black holes. The consequent restriction of the quantum  observables
${\hat h}_{\alphaa}, {\hat g}_{\alphaa (r)}$ to loops $\alpha^k$ of length at most $s$ leads to the 
determination of $\epsilon$ via equation (\ref{2.19}). Moreover, the derivations of various  bounds in this 
work lead us to believe that (i) it is only for loops with length  close to $s$  that the fluctuations
of the linearised observables can be  close to $\epsilon$
\footnote{ More work is necessary to establish exactly which $\alpha^k$ yield $\Delta g_{\alphaa (r)}\approx \epsilon$
to nontrivial leading order in $\epsilon$. However the bound (\ref{2.19a}) clearly indicates that any such $\alpha^k$ should
have lengths $\approx$ s.}
 (ii) the seminorms of the various distributions 
encountered are determined primarily by the action of these distributions on states of type  (b) in section 4B
with $\eta_I, \beta_J$  of length $\approx s$ and $M,P\approx 1$. 
Thus, $s$ seems to play a key role in characterising the nature and content of the approximate descriptions 
at both the classical and the quantum levels.

A second feature of our work is (for reasons described in section 3A (ii)) the use of distributions as opposed to 
kinematically normalizable states. Since we believe that distributional states will play a key role in future
developments, it would be profitable to explore structures on $\Phi^*_{kin}$. In the theory of
of infinite dimensional vector spaces, 
topologies are typically defined via (families of) seminorms \cite{yamasaki}. 
In analogy with this and in view of the fact that the seminorm $||\;||^{\epsilon}$ on $\Phi^*_{kin}$
has played a key role in our considerations, we advocate further study of seminorms on the space $\Phi^*_{kin}$.

The seminorm $||\;||^{\epsilon}$ provided us with the beginnings of a perturbative treatment of the 
constraints in terms of the expansion parameter $\epsilon$. The first step was to 
verify that  the linearised constraints mapped our states  into the kernel of $\me$. In showing this
we did not display an explicit regularization of the constraints. Rather, we predicated our arguments
on certain expectations of any ``reasonable'' regularization procedure.The discussion in section 6 indicates
that any such reasonable regularization procedure must provide some preferred role to the weave $\Delta$.
This seems to be in contrast to the standard Thiemann regularization \cite{qsd} of say, the full Hamiltonian
constraint, in which there is no preferred background structure. We believe that this possible conflict between the 
(putative) perturbative treatment of the constraints and the available nonperturbative treatment is of a generality 
which transcends the details of our particular constructions. Further work is needed to verify if such a
conflict is indeed present in our work and if so, to seek a resolution.
Another important open issue is to understand better the relation between the $U(1)^3$ gauge transformations
in linearised theory and the $SU(2)$ ones of LQG. The discussion in section 7 was indicative of 
our lack of understanding of this relation. We feel that a first exploratory step towards more clarity
would be  
to see if we can use a more sophisticated seminorm in accordance with (i)-(ii)
at the end of  section 7. In this regard, it would be worthwhile to base our discussion of $|\phi >$ in
(\ref{seminorm1}) in terms of graph-based spinnets rather than loop based wave functions.

The key issue underlying the above paragraph is that of the relation between (putative) perturbative 
solutions to the constraints and the known structure of the nonperturbative ones as 
spatially diffeomorphism invariant, $SU(2)$ gauge invariant distributions \cite{alm2t}. As mentioned
earlier in this work, we have merely shown the existence of a map $\me$ satisfying the various requirements of section 3.
While this was a nontrivial and necesssary exercise,  it is important to emphasize that our constructions were based 
primarily on {\em mathematical} analogies between  $U(1)^3$ and $SU(2)$ structures. Without additional physical
insight into the significance of the various choices we have made, we are unsure if these choices (or suitable modifications
thereof) ensure 
that (putative) perturbative solutions converge to nonperturbative ones (or,indeed, if they ensure even the 
existence of a well defined perturbative scheme). Since the issue is a deep and quite
general one, clearly more work and thought is needed for further progress. Indeed, the main virtue of our work
may be that it offers a concrete and detailed context wherein this issue may be analysed.

It would be of interest to see if our considerations  can be generalised to construct LQG correspondents
of states other than the linearised theory vacuum. It would also be interesting to adapt  our general
framework so as to find LQG correspondents of other known exactly solvable restrictions of full gravity 
such as Einstein- Rosen waves \cite{kuchar}. From equation (\ref{3.8}) it is tempting to speculate
that such `consistency'  with a multitude of exactly solvable restrictions of full gravity 
may lead to information about the scalar product of full quantum gravity.

There have been other efforts to construct the vacuum state in LQG. We comment briefly on three of these,
namely \cite{conrady}, \cite{ttcomplexifier} and \cite{aavac}. The recent work \cite{conrady} is a very interesting
attempt to construct the vacuum state as a kinematically normalizable state in LQG rather than, as in this work,
a genuine distribution. Due to our unfamiliarity with the detailed considerations of \cite{conrady}, we are unable to
comment further on its relation to our work.  The work \cite{ttcomplexifier} contains a nice description
of the author's  beautiful complexifier  construction of coherent states. In the LQG context this construction yields
genuinely distributional states. However, the author's viewpoint is that since such states do not seem to 
support the action of flux operators \cite{ttcomplexifier}, it is appropriate to use kinematically normalizable 
`cut off' versions 
of these states which are similar to the the gauge theory coherent states of \cite{ttvac}. As mentioned earlier,
our viewpoint is that distributions {\em are} the appropriate structures. It would be interesting to see if the 
complexifier generated distributions satisfy the requirements of section 3B in the context of some suitably defined 
$\me$. Distributional states are also used in \cite{aavac}. Our work seems to be closest in spirit to \cite{aavac}
and it may be of interest to draw parallels between the graphs chosen as `probes' in 
\cite{aavac} and the states in section 4B (b) which go into the definition of the seminorm $||\;||^{\epsilon}$.

In closing, we reiterate that many of the considerations of this work were motivated by the earlier work
of Iwasaki and Rovelli \cite{IR}. Since the work of Zegwaard \cite{joost} also tried to address the same
problem as \cite{IR}, it would be worthwhile to revisit it in the light of our work here.

\noindent {\bf Acknowledgements}: We thank Carlo Rovelli and Thomas Thiemann for encouragement. We are indebted to
Daniel Sudarsky and Alejandro Perez for their critiscisms. We thank Abhay Ashtekar for encouragement and
crucial insights as well as Chris Van Den Broeck for numerous discussions. We are grateful for the hospitality of the 
Center for Gravitational Physics and Geometry, PSU, where we discussed a preliminary version of this work.
A significant part of this work was completed when the author was visiting the Ludwig Maximillian University
in Munich. We thank Professor Mukhanov  of LMU for his kind support during this period. We thank Supurna Sinha
for pointing us to the Poisson Sum Formula. We are indebted to  Alok Laddha for his comments and for a careful
perusal of a draft version of this work.

\appendix

\section{}
We sketch the main steps involved in showing (\ref{Eweave}). Let $\psi_{\alpha}$ be any spinnet
based on a graph $\alpha$ with edges $e_I, I=1..N$ (we shall use the notation $|\psi_{\alpha}>$
interchangeably with $\psi_{\alpha}$).  Then (see for example equation (3.6) of \cite{area}) we 
have that
\begin{eqnarray}
({\hat E}^a_{(r)i}({\vec x}) &\Psi_{weave})[\psi_{\alpha}]&
=: \Psi_{weave}[{\hat E}^a_{(r)i}({\vec x})\psi_{\alpha}]\nonumber\\
&= il_P^2\gamma &\sum_I 
\int dt{\dot e}_I^a
{e^{-{|{\vec x}- {\vec e}_I(t)|^2 \over 2 r^2}}\over (2\pi r^2)^{3\over 2}}
\Psi_{weave} [(h_{e_I}(1,t)\tau_ih_{e_I}(t,0))_A^B{\partial \psi_{\alpha}\over \partial h_{e_I}(1,0)_A^B}].
\label{ehatpsi}
\end{eqnarray}
Here $h_{e_I}(v,u)$ is the holonomy along the $I$th edge from parameter value $u$ to parameter
value $v$, the beginning and end points of $e_I$ being at parameter value 0 and 1. Note that the 
argument of $\Psi_{weave}$ above is obtained from $\psi_{\alpha}$ by inserting 
the intertwiner $\tau_i$  at the point ${\vec e}_{I}(t)$ on the $I$th edge of $\alpha$. Since
$\Psi_{weave} \in \Phi^*_{kin}$, we have that 
$\Psi_{weave} [(h_{e_I}(1,t)\tau_ih_{e_I}(t,0))_A^B{\partial \psi_{\alpha}\over \partial h_{e_I}(1,0)_A^B}]$
is finite. Hence we can neglect contributions to the integral in (\ref{ehatpsi})coming from any 
sets of measure zero in parameter space. In particular, we can drop contributions from
pre-existing vertices ${\vec e}_I(0), {\vec e}_I(1)$.

\noindent {\bf Lemma 1}: If $\psi_{\alpha}$ is orthogonal to all $|\Delta, \{{\vec x}_k\}>$, then 
$({\hat E}^a_{(r)i}({\vec x}) \Psi_{weave})[\psi_{\alpha}]=0$.

\noindent Proof: Note that $\psi_{\alpha}$ is orthogonal to $|\Delta, \{{\vec x}_k\}>$ iff 
any of the following are true:\\
\noindent (i) $\alpha$ is different from $\Delta$,\\
\noindent (ii)$\alpha = \Delta$ but has at least one edge with $j\neq {1\over 2}$, \\
\noindent (iii) $\alpha =\Delta$, every edge is labelled by $j= {1\over 2}$ and there
exists at least one point in $\Delta$ at which the intertwiners for $\psi_{\alpha}$ and
$|\Delta, \{{\vec x}_k\}>$ are orthogonal i.e.
$\sum_{i_1..i_t,j_1..j_s} C_{(\alpha )j_1..j_s}^{i_1..i_t} 
{C^*_{(\Delta, \{{\vec x}_k\})}}^{i_1..i_t}_{j_1..j_s}=0$. 
Here $C_{(\alpha )}$ is the intertwiner for the vertex in question for
$\psi_{\alpha}$, ${C_{(\Delta, \{{\vec x}_k\})}}$ is the intertwiner
for the same vertex for $|\Delta, \{{\vec x}_k\}>$ and $i_1..i_t$ label
the representations on the 
incoming edges (for which the vertex is a `target') and $j_1..j_s$
label the representations on the outgoing edges (for which the vertex is 
a `source'). Note that these intertwiners may be further split into a 
product of invariant intetwiners and representation vectors  but this is not 
necessary for our purposes. 

The result (iii) is obtained straightforwardly from the definition of the 
intertwiners and the orthogonality properties of the representations of 
$SU(2)$ with respect to the Haar measure. Clearly if any of (i)-(iii) are
true of $\psi_{\alpha}$, the same is true of 
$(h_{e_I}(1,t)\tau_ih_{e_I}(t,0))_A^B{\partial \psi_{\alpha}\over \partial h_{e_I}(1,0)_A^B}$
except at most on sets of measure zero in parameter space.
Thus, for $\psi_{\alpha}$ orthogonal to all 
 $|\Delta, \{{\vec x}_k\}>$, we have that 
\begin{equation}
({\hat E}^a_{(r)i}({\vec x}) \Psi_{weave})[\psi_{\alpha}]=0= \delta^a_i \Psi_{weave})[\psi_{\alpha}].
\label{Eweave1.1}
\end{equation}

\noindent {\bf Lemma 2}: Let the extent $L$ of $\Delta $ be 
such that $s_{curv} >> L>> r >>l_P$. Further, let $L>>s$ and let ${\vec x}$ lie in the 
probe region of size $(Ss)^3$ about the origin (see 3A(i) for the definition of $S$) so that
$L>>|{\vec x}|$. Then 
\begin{equation}
({\hat E}^a_{i(r)}({\vec x})\Psi_{weave})[|\Delta, \{{\vec x}_k\}>] = 
(1+ O^{\infty}(\delta )+O^{\infty}(\tau ) )\delta^a_i \Psi_{weave})[|\Delta, \{{\vec x}_k\}>]
\label{Eweave2}
\end{equation}
with $\delta = {\sqrt{l_P^2\gamma\over 2}\over r}$ and 
$\tau = {r\over L}$.

\noindent Proof: From (\ref{ehatpsi}) and the definition of $\Psi_{weave}$, we have that 
$({\hat E}^a_{i(r)}({\vec x})\Psi_{weave})[|\Delta, \{{\vec x}_k\}>]=0$ if $a\neq i$.
Hence the result must be proportional to $\delta^a_i$. Let us suppose that $a=i=1$
(our considerations below trivially extend to any other choice of $a=i$). Then from 
equations (\ref{ehatpsi})  and (\ref{defweave}), it is straightforward to show that
\begin{eqnarray}
\Psi_{weave} [{\hat E}^a_{(r)i}({\vec x})|\Delta, \{{\vec x}_k\}>]
&=& (-i)^N(\sum_{L_z}{\sqrt{l_P^2\gamma\over 2}}{e^{-{(z-L_z)^2\over 2r^2}}\over {\sqrt{2\pi}{ r}}})\nonumber\\
& &(\sum_{L_y}{\sqrt{l_P^2\gamma\over 2}}{e^{-{(y-L_y)^2\over 2r^2}}\over {\sqrt{2\pi} {r}}})
\int_{-L}^{L} d{\bar x} {e^{-{(x-{\bar x})^2\over 2r^2}}\over {\sqrt{2\pi}{ r}}} .
\label{gaussian}
\end{eqnarray}
Here $(L_y,L_z)$ range over the lattice points in the $x=0$ plane. Note that from (\ref{equant}),
the condition $s_{curv}>>L$ is the same as $r^2>>l_PL$. Straightforward estimates 
using the latter condition and the condition that $L>>|{\vec x}|$, in conjunction 
with the Poisson Summation formula 
\footnote{ We are indebted to Supurna Sinha for pointing us to the Poisson Sum formula
which plays this  key role as an estimation technique in our work.}
\cite{poisson} yield
\begin{equation}
(\sum_{L_z}{\sqrt{l_P^2\gamma\over 2}}{e^{-{(z-L_z)^2\over 2r^2}}\over {\sqrt{2\pi}{ r}}}),
(\sum_{L_y}{\sqrt{l_P^2\gamma\over 2}}{e^{-{(y-L_y)^2\over 2r^2}}\over {\sqrt{2\pi}{ r}}}) 
= 1 +O^{\infty}(\delta )+ O^{\infty}(\tau ).
\end{equation}
Also it is easy to check that 
\begin{equation}
\int_{-L}^{L} 
d{\bar x} {e^{-{(x-{\bar x})^2\over 2r^2}}\over {\sqrt{2\pi}{ r}}} = 1 +O^{\infty}(\tau ).
\end{equation} 
This completes the proof of Lemma 2.

In order to obtain equation (\ref{Eweave}) from Lemma 2, it suffices to choose 
$\delta = \epsilon^{\alpha},\; \tau= \epsilon^{\beta}, \;\alpha,\beta>0$. Below we display 
2 explicit choices of $\epsilon$ dependence of the various scales involved which
satisfy all our  requirements.

\noindent(a) Choose $s$ independent of $\epsilon$, $r= \epsilon^{-{1\over 2}}$,
$s_{curv} = \epsilon^{-1} s$, $L = \epsilon^{-{3\over 4}}s$.  Thus $\alpha ={1\over 2}$
and $\beta={1\over 4}$ and we have that $s_{curv}>>L>>r>>s \sim O(1)$.

\noindent (b) $s$ dependent on $\epsilon$, $s\neq 0$ as $\epsilon \rightarrow 0$.
Choose $\alpha >{1\over 2}$ and $\beta >0, \; \alpha >\beta >\alpha -1$.
Let $s= \epsilon^{1-2\alpha}l_P$, $r= \epsilon^{-\alpha}l_P$, $s_{curv} =\epsilon^{-2\alpha}l_P$,
$L= \epsilon^{-(\alpha +\beta )}l_P$.
Then for $\alpha >1$, we have that $s_{curv}>>L>>s>>r$ and 
for $\alpha <1$, $s_{curv} >> L >>r >>s $ .

As mentioned at the end of section 2, we shall set $s>>r$ i.e. we shall restrict attention
to the choice (b) with $\alpha >1$.

\section{}

In this appendix we prove the identities (\ref{gnweave1}),(\ref{gnweave2}) under the assumption that 
(\ref{hnweave}) holds for all $\vec x$ on the spatial slice. 
From (\ref{gab}) we have that 
\begin{equation}
G_{\alphaa(r)ab}({\vec x})
= 
(1+2i\gamma )g^{(+)}_{\alphaa (r)ab}({\vec x}) -(1-2i\gamma )g^{(-)}_{\alphaa(r)ab}({\vec x})
\label{gabx}
\end{equation}
where $g^{\pm}_{\alphaa(r)ab}$ are the positive/negative helicity components of $g_{\alphaa(r)ab}$,
\begin{eqnarray}
g^{ab}_{\alphaa(r) }({\vec x})&:=&\int {d^3k\over (2\pi)^{3\over 2}}e^{i{\vec k}\cdot{\vec x}}k X^{ab}_{\alphaa (r)}({\vec k})
\nonumber \\
&=& \oint_{\alpha^b} dt{{\dot{\alpha^b}}^a\over (2\pi)^{3\over 2}}I ({\vec x}- {\vec \alpha^b}(t)).
\label{gI}
\end{eqnarray}
$I({\vec y})$ is defined as
\begin{eqnarray}
I({\vec y})&:=& \int {d^3k\over (2\pi)^{3\over 2}}e^{i{\vec k}\cdot{\vec y}}ke^{-k^2r^2\over 2}
\nonumber \\
&=& {4\pi\over (2\pi)^{3\over 2}r^3y}\int_0^{\infty}du \sin({y\over r}u) u^2 e^{-u^2\over 2}
\label{defI}
\end{eqnarray}
where we have denoted $|{\vec y}|$ by $y$.

From (\ref{hnweave}),(\ref{GEr}) and (\ref{gabx})
it follows that for $|\psi_{\Delta}>= |\Delta >$ or $|\psi_{\Delta}>= |\Delta \{{\vec x}_k\} >$
\begin{eqnarray}
|\prod_{j=0}^{n-1} {\hat G}^{\dagger}_{\alphaa_{n-j}}\Psi_{weave} [|\psi_{\Delta} >]|
&\leq & O^{\infty}(\epsilon )
(\prod_{i=1}^n(\int d^3x_i |G^{*a_ib_i}_{\alphaa_i(r)}({\vec x}_i)G_{\alphaa_i(r)a_ib_i}({\vec x}_i)|^{1\over 2}) \nonumber\\
&\leq & O^{\infty}(\epsilon )\prod_{i=1}^n(\int d^3x_i\sum_{a_i,b_i=1..3}(1+4\gamma^2)^{1\over 2}
|g_{\alphaa_i(r)a_ib_i}({\vec x}_i)| .
\label{gweaveabs}
\end{eqnarray}
We estimate a bound on $|g_{\alphaa(r)ab}({\vec x})|$ through an analysis of its asymptotic behaviour as 
follows. From \cite{bateman1} we have that 
\begin{equation}
\int_0^{\infty}du \sin({y\over r}u) u^2 e^{-u^2\over 2} = 
{2y\over r}e^{-y^2\over 2r^2}{}_1F_1(-{1\over 2};{3\over 2};{y^2\over 2r^2}),
\label{1f1}
\end{equation}
where ${}_1F_1(a;c;z)$ is the confluent hypergeometric function. From \cite{bateman2} the 
asymptotic 
behaviour of this function as $z\rightarrow \infty$ is 
\begin{equation}
{}_1F_1(a;c;z) ={\Gamma (c) \over \Gamma (a)}e^z z^{a-c}(1+ O({1\over z})).
\label{asymp1f1}
\end{equation}
The parameters $z,a,c$ in (\ref{asymp1f1}) are identified through  (\ref{1f1}),(\ref{gI}) and 
(\ref{defI})  as $z={|{\vec x}-{\vec \alpha_b}(t)|^2\over 2r^2}$,
$a=-{1\over 2}$ and $c= {3\over 2}$.
Using (\ref{asymp1f1}) in (\ref{gI}) and (\ref{defI}), it is straightforward to obtain the bound
\begin{equation}
|g_{\alphaa(r)ab}({\vec x})| < {2s\over \pi^2 {\rm min}_t |{\vec x}-{\vec\alpha^b}(t)|^4}
\label{boundg1}
\end{equation}
where ${\rm min}_t f(t)$ refers to the minimum value of the function $f(t)$ over the entire loop $\alpha_b$ and 
the bound is valid for all ${|{\vec x}-{\vec\alpha^b}(t)|\over r} > \lambda$ for sufficiently large $\lambda$.
Since
the triplets of loops $\alphaa_i, i=1..n$ are all confined to a region of size $Ss$  about the origin (see section 3A(i)),
we have that ${\rm min}_t |{\vec x}-{\vec\alpha^b}(t)| > |{\vec x}| -Ss$.

Then it follows from (\ref{boundg1}) that for $|{\vec x}| > r\lambda + 2Ss$ that
\begin{equation}
|g_{\alphaa(r)ab}({\vec x})| <{32s\over \pi^2  |{\vec x}|^4}.
\label{boundg2}
\end{equation}
Also note that from (\ref{gI}) and (\ref{defI}) the following bound is easily obtained:
\begin{equation}
\int _{|{\vec x}|<r\lambda +2Ss}d^3x |g_{\alphaa (r)ab}({\vec x})| \leq (\lambda +{2Ss\over r})^3 {s\over \pi r}.
\label{boundg3}
\end{equation}
From (\ref{boundg2}),(\ref{boundg3}) it follows that 
\begin{equation}
\sum_{a,b=1..3}\int d^3x |g_{\alphaa (r)ab}({\vec x})| \leq 
{9s\over \pi r}\{(\lambda +{2Ss\over r})^3 + {128\over (\lambda +{2Ss\over r})}\}.
\label{boundg4}
\end{equation}
Using the above bound in (\ref{gweaveabs}) with either (a) or (b) of Appendix A and  the 
$\epsilon$- independence of  
$S, \lambda, n$,  we obtain the desired result, namely that
\begin{equation}
|\prod_{j=0}^{n-1} {\hat G}^{\dagger}_{\alphaa_{n-j}}\Psi_{weave} [|\psi_{\Delta} >]| \leq O^{\infty}(\epsilon ).
\end{equation}

\section{}

Let  
\begin{equation}
O^H_I(A)\in \{{{\rm Tr}H_{\eta}\over 2},{\rm Tr}H_{\eta}\tau^k,\;\;\eta \in G^s_{\Delta loop}, k=1..3\} ,\;\; I=1..N.
\label{defoh}
\end{equation}
Thus each $O^H_I(A)$ is  (half) the trace of a holonomy or the trace of the product of a holonomy with $\tau^i, i\in\{1,2,3\}$,
the holonomy being around any loop of length at most equal to $s$.
Then the following bounds hold.
\begin{eqnarray}
\left| \left(\prod_{j=0}^{n-1}\int d^3k_{n-j}(G_{\alphaa_{n-j}(r)a_{n-j}b_{n-j}}({\vec k}_{n-j}))^*
{\hat E}^{a_{n-j}b_{n-j}}_{(r)}({\vec k}_{n-j})\prod_{I=1}^N O^H_I(A)\right)_{A=0}\right|:= &  \nonumber \\
\left| \left(\int d^3k_{n}(G_{\alphaa_{n}(r)a_{n}b_{n}}({\vec k}_{n}))^*
{\hat E}^{a_{n}b_{n}}_{(r)}({\vec k}_{n})..\int d^3k_{1}(G_{\alphaa_{1}(r)a_{1}b_{1}}({\vec k}_{1}))^*
{\hat E}^{a_{1}b_{1}}_{(r)}({\vec k}_{1})\prod_{I=1}^N O^H_I(A)\right)_{A=0}\right| &\nonumber \\
\leq N^n \epsilon^{2n} c_n,  &
\label{4}
\end{eqnarray}
with $c_n= ({27\sqrt{1+4\gamma^2}\gamma \over 4\pi^2})^n$.
\begin{eqnarray}
&| \int {d^3k\over 2i\gamma l_P^2}k[(1\mp 2i\gamma ){\hat E}^{\pm}_{(r)}({\vec k}) {\hat E}^{\pm\dagger}_{(r)}({\vec k})]&\nonumber\\
&\prod_{j=0}^{n-1}\int d^3k_{n-j}(G_{\alphaa_{n-j}(r)a_{n-j}b_{n-j}}({\vec k}_{n-j}))^*
{\hat E}^{a_{n-j}b_{n-j}}_{(r)}({\vec k}_{n-j})\prod_{I=1}^N O^H_I(A))_{A=0}|&  \nonumber \\
&\leq N^{n+2} \epsilon^{2n+2}d_n, \;\; d_n={3\over \sqrt{2}}({27\sqrt{1+4\gamma^2}\gamma \over 4\pi^2})^{n+1} .& 
\label{5}
\end{eqnarray}
Here $\alphaa_i, i=1..n$ are $n$ triplets of loops, each loop with length at most equal to $s$, and ${\hat E}^{\pm}$
are the positive and negative helicity components of the triad operator.

\noindent Sketch of the proof: We shall only describe the main steps of the proof. The details
are straightforward to work out and the interested reader may easily do so.

\noindent (i) The following bound holds
\begin{equation}
\left| \left(\prod_{j=0}^{n-1}
{\hat E}^{a_{n-j}b_{n-j}}_{(r)}({\vec k}_{n-j})O^H_I(A)\right)_{A=0}\right| \leq 
({l_P^2\gamma s\over (2\pi)^{3\over 2}})^n\prod_{i=1}^n e^{{-k_i^2r^2\over 2}}.
\label{1}
\end{equation}
Note that the action of the smeared triad operator ${\hat E}^{a}_{(r)i}({\vec k})$ on a cylindrical function 
$\psi_{\alpha}$
based on a graph $\alpha$ with 
edges $e_J$ is 
\begin{equation}
{\hat E}^a_{(r)i}({\vec k}) \psi_{\alpha}
=il_P^2\gamma \sum_{J}
\int dt{\dot e}_J^a
{e^{-i{\vec k}\cdot {\vec e}_J(t)}e^{-k^2r^2 \over 2}\over (2\pi )^{3\over 2}}
\{(h_{e_J}(1,t)\tau_ih_{e_J}(t,0))_A^B{\partial \psi_{\alpha}\over \partial h_{e_J}(1,0)_A^B}\}.
\label{ehatpsi1}
\end{equation}
For $\psi_{\alpha}$ of the type $O^H_I$, the term in curly brackets corresponds to an insertion of 
$\tau^i$ at parameter value $t$. The action of a product of $n$ factors of triad operators on $O^H_I$
yields $n$ integrations over the loop labelling $O^H_I$, with $n$ insertions of $\tau$ matrices along the 
loop. The evaluation at $A=0$ of $O^H_I(A)$ modified by these insertions yields the trace of a product of $\tau$
matrices.
This trace is bounded by unity.  The $n$ integrations along the loop are responsible for the $s^n$ factor in the bound
(\ref{1}) and the origin of the $({l_P^2\gamma \over (2\pi)^{3\over 2}})e^{{-k_i^2r^2\over 2}}$ factors 
is obvious from (\ref{ehatpsi1}).

\noindent (ii) The following bound holds.
\begin{equation}
\left| \left(\prod_{j=0}^{n-1}
{\hat E}^{a_{n-j}b_{n-j}}_{(r)}({\vec k}_{n-j})\prod_{I=1}^N O_I(A)\right)_{A=0}\right| \leq 
N^n({l_P^2\gamma s\over (2\pi)^{3\over 2}})^n\prod_{i=1}^n e^{{-k_i^2r^2\over 2}}.
\label{2}
\end{equation}
This follows from a straightforward estimate of the number of terms of type (\ref{1}).

\noindent (iii) For any `smearing' function $G_{a_1..a_nb_1..b_n}({\vec k}_1,..,{\vec k}_n)$ the following 
bound holds.
\begin{eqnarray}
&\left| \left(\int \prod_{i=1}^nd^3k_i(G_{a_1..a_nb_1..b_n}({\vec k}_1,..,{\vec k}_n))^*
\prod_{j=0}^{n-1}
{\hat E}^{a_{n-j}b_{n-j}}_{(r)}({\vec k}_{n-j})\prod_{I=1}^N O_I(A)\right)_{A=0}\right|&\nonumber \\
&\leq 
N^n({9l_P^2\gamma s\over (2\pi)^{3\over 2}})^n\int \prod_{i=1}^n d^3k_ie^{{-k_i^2r^2\over 2}}
(\sum_{\{a_i,b_i\}}|G_{a_1..a_nb_1..b_n}({\vec k}_1,..,{\vec k}_n)|^2)^{1\over 2}.&
\label{3}
\end{eqnarray}  
To see this, note that the left hand side of the above equation 
 is bounded by the integral of the product of the norms
of the smearing function and the quantity $\left(\prod_{j=0}^{n-1}
{\hat E}^{a_{n-j}b_{n-j}}_{(r)}({\vec k}_{n-j})\prod_{I=1}^N O_I(A)\right)_{A=0}$.  The norm of the latter is 
bounded
by the right hand side of (\ref{2}) augmented by a factor of $9^n$ 
which comes fron the sum over the indices $a_i,b_i$. As a result, equation (\ref{3}) follows.

\noindent (iv) Set $G_{a_1..a_nb_1..b_n}({\vec k}_1,..,{\vec k}_n)= \prod_{i=1}^n G_{\alphaa_i (r)a_i b_i} ({\vec k}_i)$
in (\ref{3}). Using (\ref{gab}), it is easily verfied that 
\begin{equation}
(\sum_{a,b}|G_{\alphaa_i (r)a b} ({\vec k}_i)|^2)^{1\over 2} \leq  k_i e^{-k_i^2r^2\over 2}(1+4\gamma^2)^{1\over 2}
                                              {3s\over (2\pi)^{3\over 2}}.
\end{equation}
The integral over each ${\vec k}_i$ in (\ref{3}) furnishes a factor of ${2\pi\over r^4}$. This, together 
with $\epsilon = {l_Ps\over r^2}$ from (\ref{equant}) yields (\ref{4}).
Similarly, for an  appropriate choice of $G_{a_1..a_nb_1..b_n}({\vec k}_1,..,{\vec k}_n)$ and straightforwardly
derived bounds thereon, equation (\ref{5}) may also be obtained.

\section{}

We prove the identities (\ref{oeoe1}) through the following steps.

\noindent (1) 
Consider a set of operators ${\hat O}_i, i= 0..n$ such
that each ${\hat O}_i$ is of the type defined in equation (\ref{defO}).
Denote the 
operators ${\hat E}^{a_j}_{(r)i_j}$ by ${\hat E}_j$ and the Kronecker deltas $\delta^{a_j}_{i_j}$ by 
$\delta_j$. Define ${\hat O}_{1,2..,n}$ and ${\hat O}_{1,2,..,\not{i},.\not{j},.,n}$
by 
\begin{eqnarray}
{\hat O}_{1,2..,n} &:=& {\hat O}_1{\hat O}_2...{\hat O}_n \\
{\hat O}_{1,2,..,\not{i},.\not{j},.,n} &:=&
{\hat O}_1{\hat O}_2..{\hat O}_{i-1}
[{\hat E}_i,{\hat O}_i {\hat O}_{i+1}..{\hat O}_{j-1}[{\hat E}_j {\hat O}_j...{\hat O}_n]..] .
\label{notindex}
\end{eqnarray}
Thus, in definition (\ref{notindex}), each `slashed' index implies a commutator of the string of operators
which  follow the index with the triad operator labelled by that index. Then the following lemma holds.

\noindent {\bf Lemma 3}: 
\begin{eqnarray}
\prod_{i=1}^n({\hat E}_i-\delta_i){\hat O}_i &=& (E_1 -\delta_1){\hat O}_1 (E_2 -\delta_2){\hat O}_2..(E_n -\delta_n){\hat O}_n
\nonumber \\
&=& {\hat O}_{1..n} \prod_{i=1}^n(E_i -\delta_i)  + \sum_{I=1}^n(\sum_{i_1,i_2..,i_I}{\hat O}_{1,2,..,\not{i_1},.\not{i_I},.,n}
                                                          \prod_{j\not\in \{i_1,..,i_I\}}(E_j-\delta_j),
\label{lemma}
\end{eqnarray}
where the second sum in the second term on the right hand side  is over all choices of `slashed indices' 
$i_1 < i_2..<i_I$.

\noindent Proof (by induction):
Equation (\ref{lemma}) is easily verified for $n=1$. Let it be true for $n=m$. For $n=m+1$, the left hand side
of (\ref{lemma}) is $\prod_{i=1}^{m+1} ({\hat E}_i- \delta_i){\hat O}_i$.. We apply (\ref{lemma}) for $n=m$ to the 
string of operators $\prod_{i=2 }^{m+1} ({\hat E}_i- \delta_i){\hat O}_i$ to obtain
\begin{eqnarray}
\prod_{i=1}^{m+1} ({\hat E}_i- \delta_i){\hat O}_i &=&
\{({\hat E}_1 -\delta_1) {\hat O}_{1...m+1}\prod_{i=2}^{m+1}({\hat E}_i- \delta_i)\} \nonumber \\
&+&\{({\hat E}_1- \delta_1) {\hat O}_1\sum_{I=2}^{m+1}\sum_{i_2,..i_I\neq 1}{\hat O}_{2,3,..,\not{i_2}..,\not{i_I},..m+1}
                                             \prod_{j\not\in\{1,i_2..i_I\}}({\hat E}_j-\delta_j)\}  \\
&=&\{{\hat O}_{1...m+1}\prod_{i=1}^{m+1}({\hat E}_i- \delta_i) 
+ {\hat O}_{\not{1},2,...m+1} \prod_{i=2}^{m+1}({\hat E}_i- \delta_i)\} \nonumber \\
&+&\{\sum_{I=2}^{m+1}\sum_{i_2,..i_I\neq 1}{\hat O}_{\not{1},2,3,..,\not{i_2}..,\not{i_I},..m+1}
                                             \prod_{j\not\in\{1,i_2..i_I\}}({\hat E}_j-\delta_j) \nonumber \\
&+&\sum_{I=2}^{m+1}\sum_{i_2,..i_I\neq 1}{\hat O}_{1,2,3,..,\not{i_2}..,\not{i_I},..m+1}
                                             (E_1-\delta_1)\prod_{j\not\in\{1,i_2..i_I\}}({\hat E}_j-\delta_j)\} \\
&=& {\hat O}_{1...m+1}\prod_{i=1}^{m+1}({\hat E}_i- \delta_i) + 
\sum_{I=1}^{m+1}\sum_{i_1,i_2..,i_I}{\hat O}_{1,2,..,\not{i_1},.\not{i_I},.,m+1}
                                                          \prod_{j\not\in \{i_1,..,i_I\}}(E_j-\delta_j).
\end{eqnarray}
This completes the proof.

\noindent Note:
In the equations above it is understood that any set of indices $\{i_k\}$ is such that if $p<q$, then  $i_p<i_q$
and that the set of indices range over all appropriate subsets of $\{1,2,..m+1\}$ modulo  explicitly forbidden
values (for e.g. in some cases the indices have to be all different from 1).

\noindent (2)
In the connection representation the commutator $[{\hat E}_{i_1},{\hat O}_{i_1, {i_1+1},..\not{j_1}..\not{j_2}..n}]$
is a `multiplication'  operator whose action on any state $\psi ({\bar A})$ is given by 
\begin{eqnarray}
[{\hat E}_{i_1},{\hat O}_{i_1, {i_1+1},..\not{j_1}..\not{j_2}..n}]\psi ({\bar A})
&=&  \int d^3y{e^{-{|{\vec x}_{i_1}- {\vec y}|^2 \over 2 r^2}}\over (2\pi r^2)^{3\over 2}}\nonumber \\
&& (-il_P^2\gamma {\delta \over \delta A_{a}^i({\vec y})}{\hat O}_{i_1, {i_1+1},..\not{j_1}..\not{j_2}..n}(A)|_{A={\bar A}})
\psi ({\bar A})
\label{e,o}
\end{eqnarray}
with the term multiplying $\psi ({\bar A})$ on the right
 hand side of the above equation completely determined by the holonomies of the generalised connection
${\bar A}$ along graphs contained in $G_{\Delta}$ (recall that $G_{\Delta}$ is the set of all graphs which intersect
$\Delta$ in at most a finite number of points).

Using (\ref{e,o}) and Lemma 3 above, in conjunction with (a)-(c) of section 4C, it is straightforward to 
see that for $|\psi_{\Delta} > = |\Delta >$ or  $|\psi_{\Delta} > = |\Delta \{x_k\} >$, we have
\begin{eqnarray}
&\Psi^{\epsilon}_0 [{\hat O}_0(E_1 -\delta_1){\hat O}_1 (E_2 -\delta_2){\hat O}_2..(E_n -\delta_n){\hat O}_n|\psi_{\Delta} >]
=\Psi_{weave}[|\psi_{\Delta}>]
\Phi^{\epsilon}_0[{\hat O}_0{\hat O}_{{\not 1}{\not 2}..{\not n}}|\cdot >]&  \nonumber \\
&+ \sum_{I=1}^{n-1}(\sum_{i_1,i_2..,i_I}
\Psi_{weave}[\prod_{j\not\in \{i_1,..,i_I\}}(E_j-\delta_j)|\psi_{\Delta}>]
\Phi^{\epsilon}_0[{\hat O}_0{\hat O}_{1,2,..,\not{i_1},.\not{i_I},.,n}|\cdot >] & \nonumber \\
&+ \Psi_{weave}[\prod_{i=1}^n(E_i -\delta_i)|\psi_{\Delta}>]
\Phi^{\epsilon}_0[{\hat O}_0{\hat O}_{1..n}|\cdot >]& .
\label{oeoe}
\end{eqnarray}
Note that 
\begin{equation}
{\hat O}_0{\hat O}_{1,2,..,\not{i_1},.\not{i_I},.,n} |\cdot>= {\hat O}_0{\hat O}_1{\hat O}_2..{\hat E}_{i_1}{\hat O}_{i_1}..
                   {\hat E}_{i_I}{\hat O}_{i_I}..{\hat O}_n|\cdot>.
\end{equation}
 Using this in (\ref{oeoe}) and integrating
against $\prod_{i=1}^n G^*_{\alphaa_i(r)a_ib_i}$ with $\alphaa_i$ being triplets of loops in $G^s_{\Delta loop}$, we obtain
\begin{eqnarray}
\Psi^{\epsilon}_0[{\hat O}_0 {\hat G}^{\dagger}_{\alphaa_1(r)}{\hat O}_1...{\hat G}^{\dagger}_{\alphaa_n(r)}{\hat O}_n |\psi_{\Delta} >]=
\;\;\;\;\;\;\;\;\;\;\;\;\;\;\;\;\;\;\;\;&\nonumber\\
\Psi_{weave} [|\psi_{\Delta} >] \Phi^{\epsilon}_0[ {\hat O}_0(\int G^*_{\alphaa_1(r)a_1b_1}{\hat E}^{a_1b_1}_{(r)}){\hat O}_1
..(\int G^*_{\alphaa_n(r)a_nb_n}{\hat E}^{a_1b_1}_{(r)}){\hat O}_n
|\cdot >]&  \nonumber \\
+ \sum_{I=1}^{n-1}(\sum_{i_1,i_2..,i_I}
\Psi_{weave}[\prod_{j\not\in \{i_1,..,i_I\}}
{\hat G}^{\dagger}_{\alphaa_{i_j}(r)}|\psi_{\Delta}>]\;\;\;\;\;\;\;\;\;\;\;\;\;\;\;&\nonumber \\
\Phi^{\epsilon}_0[{\hat O}_0{\hat O}_1..
(\int G^*_{\alphaa_{i_1}(r)a_{i_1}b_{i_1}}{\hat E}^{a_{i_1}b_{i_1}}_{(r)}){\hat O}_{i_1}
..(\int G^*_{\alphaa_{i_I}(r)a_{i_I}b_{i_I}}{\hat E}^{a_{i_I}b_{i_I}}_{(r)}){\hat O}_{i_I}
..{\hat O}_n
|\cdot >] )& \nonumber \\
+ \Psi_{weave}[\prod_{i=1}^n{\hat G}^{\dagger}_{\alphaa_{i}(r)}|\psi_{\Delta}>]
\Phi^{\epsilon}_0[{\hat O}_0{\hat O}_{1..n}|\cdot >]\;\;\;\;\;\;\;\;\;\;\;\;\;\;\;& .
\label{ogog}
\end{eqnarray}

\noindent (3) Set 
\begin{equation}
{\hat O}_i=\prod_{I=1}^{P_i}{\hat O}^H_{i_I}\;  {\rm for} \; i=0..n-1, \;\; 
{\hat O}_n={\hat O^{\prime}}_n\prod_{I=1}^N {\hat O}^H_I \;\;{\hat O^{\prime}}_n=\prod_{I=1}^{P_n}{\hat O}^H_{n_I}.
\label{defoion}
\end{equation}
 where 
${\hat O}^H_I,{\hat O}^H_{i_I}$ are 
quantum operators corresponding to 
classical functions of the type defined by (\ref{defoh}).
Define $P:=\sum_{i=0}^{n}P_i$ and restrict attention to $n,P$ independent of $\epsilon$.
Using (\ref{defoion}), (\ref{gnweave1}), (\ref{gnweave1}) and (\ref{4}), (\ref{5}) in conjunction
with straightforward  bounds on the number of terms of type (\ref{4}), (\ref{5}), we obtain
\begin{eqnarray}
&|\sum_{I=1}^{n-1}(\sum_{i_1,i_2..,i_I}
\Psi_{weave}[\prod_{j\not\in \{i_1,..,i_I\}}{\hat G}^{\dagger}_{\alphaa_{i_j}(r)}|\psi_{\Delta}>]& \nonumber \\
&\Phi^{\epsilon}_0[{\hat O}_0{\hat O}_1..
(\int G^*_{\alphaa_{i_1}(r)a_{i_1}b_{i_1}}{\hat E}^{a_{i_1}b_{i_1}}_{(r)}){\hat O}_{i_1}
..(\int G^*_{\alphaa_{i_I}(r)a_{i_I}b_{i_I}}{\hat E}^{a_{i_I}b_{i_I}}_{(r)}){\hat O}_{i_I}
..{\hat O}_n
|\cdot >])|&\nonumber\\ 
& \leq O^{\infty}(\epsilon ) (N+P)^{n+1},&
\label{B}
\end{eqnarray}
\begin{equation}
|\Psi_{weave}[\prod_{i=1}^n{\hat G}^{\dagger}_{\alphaa_{i}(r)}|\psi_{\Delta}>]
\Phi^{\epsilon}_0[{\hat O}_0{\hat O}_{1..n}|\cdot >]| \leq O^{\infty}(\epsilon )(1+(N+P)^2O(\epsilon^2) ).
\label{A}
\end{equation}
In the above equations the $O^{\infty}(\epsilon ),O(\epsilon^2)$ terms are indpendent of $N$.
Thus (\ref{B}) and (\ref{A}) display the $N$ dependence of the bounds explicitly.
Note that since $P_i \geq 1$, we have that $P>n$. Using this in conjunction with the above bounds
in (\ref{ogog}) we obtain
\begin{eqnarray}
&\Psi^{\epsilon}_0[{\hat O}_0 {\hat G}^{\dagger}_{\alphaa_1(r)}{\hat O}_1...{\hat G}^{\dagger}_{\alphaa_n(r)}{\hat O^{\prime}}_n
\prod_{I=1}^N{\hat O}^H_I|\psi_{\Delta} >&
\nonumber\\
&=\Psi_{weave} [|\psi_{\Delta} >] \Phi^{\epsilon}_0[ {\hat O}_0(\int G^*_{\alphaa_1(r)a_1b_1}{\hat E}^{a_1b_1}_{(r)}){\hat O}_1
..(\int G^*_{\alphaa_n(r)a_nb_n}{\hat E}^{a_nb_n}_{(r)}){\hat O^{\prime}}_n\prod_{I=1}^N{\hat O}^H_I
|\cdot >]&  \nonumber \\
& +O^{(\infty )} (\epsilon ) {(N+P)^{n+1}}.
\label{C}
\end{eqnarray}
Using the definition of the seminorm in section 4B, it is straightforward to see that 
(\ref{C}) implies that 
\begin{equation}
||{\hat O^{\prime}}_n{\hat G}_{\alphaa_n(r)}...{\hat O}_1{\hat G}_{\alphaa_1(r)}{\hat O}_0\Psi^{\epsilon}_0||^{\epsilon}
=||\Psi^{\epsilon} ||^{\epsilon} + O^{\infty}(\epsilon )
\end{equation}
where $\Psi^{\epsilon}$ is a distribution of the type defined in section 4C for which \\
$\Phi^{\epsilon} = {\hat O^{\prime}}_n(\int G_{\alphaa_n(r)a_nb_n}{\hat E}^{a_nb_n}_{(r)})..
{\hat O}_1(\int G_{\alphaa_1(r)a_1b_1}{\hat E}^{a_1b_1}_{(r)}){\hat O}_0\Phi^{\epsilon}_0$
This completes the proof of (\ref{oeoe1}).

\section{}
 Let $\beta^i$ be loops of size $r$. Let $\alpha^i= (\beta^i)^n$ (i.e. $\alpha^i$ is obtained
by going around $\beta^i$ '$n$' times) with $n= {s\over r}$. Then it follows that 
\begin{eqnarray}
(\Delta {\hat g}_{\alphaa (r)})^2 &=&
l_P^2 (1+4\gamma^2)\int d^3k k e^{-k^2r^2} (|X^+_{\alphaa}({\vec k})|^2 +
|X^-_{\alphaa}({\vec k})|^2) \\
&=& {s^2\over r^2}
l_P^2 (1+4\gamma^2)\int d^3k k e^{-k^2r^2} (|X^+_{\betaa}({\vec k})|^2 +
|X^-_{\betaa}({\vec k})|^2) \\
&=& {s^2\over r^2}{l_P^2\over r^4}(1+4\gamma^2)
\int d^3u u e^{-u^2} (|X^+_{\betaa}({{\vec u}\over r})|^2 +
|X^-_{\betaa}({{\vec u}\over r})|^2),
\label{betaunity}
\end{eqnarray}
where we have defined the dimensionless variable ${\vec u}=r{\vec k}$.
Next, note that 
\begin{equation}
X^{a}_{\beta^k}({{\vec u}\over r}) = rX^{a}_{\beta_1^k}({\vec u}),
\label{rbeta}
\end{equation}
where 
\begin{equation}
X^{a}_{\beta_1^k}({\vec u}) := {1\over (2\pi)^{3\over 2}}
\oint_{\beta_1^k}ds e^{-i{\vec u}\cdot {\vec \beta^k_1}(s)}{\dot \beta_1}^a ,
\end{equation}
with ${\vec \beta^k_1}(s):= {{\vec \beta^k}(s)\over r}$. Since $\beta^k$ is a loop of length $r$,
$\beta_1^k$ is a `unit' size loop. For generic  $\beta_1^k$  we expect that there exist
real $a,b$ and positive definite $c$, all independent of $\epsilon$ such that for 
$u\in [a,b]$, we have that
\begin{equation}
|X^{\pm}_{\betaa_1}({\vec u})| \geq c .
\label{xabetac}
\end{equation}
Using (\ref{rbeta}) and (\ref{xabetac}), we obtain 
\begin{equation}
(\Delta {\hat g}_{\alphaa (r)})^2 \geq \epsilon^2
                                (1+4\gamma^2)2c^2 \int_a^b d^3u u e^{-u^2}
\label{lowerbound}
\end{equation}

\end{document}